\documentclass[fleqn,12pt]{wlscirep}

\usepackage[utf8]{inputenc}
\usepackage[T1]{fontenc}
\usepackage[symbol]{footmisc}

\usepackage{pdfpages}

\title{Machine learning a time-local fluctuation theorem for nonequilibrium steady states}

\author[1]{Stephen Sanderson} \author[1,2]{Charlotte F. Petersen} \author[1,3,*]{Debra J. Searles}

\affil[1]{Australian Institute for
Bioengineering and Nanotechnology, The University of Queensland, Brisbane, QLD, 4072, Australia}

\affil[2]{School of Chemistry, University of Melbourne, Melbourne, Victoria, 3010, Australia}

\affil[3]{School of Chemistry and Molecular Biosciences, The University of
Queensland, Brisbane, QLD, 4072, Australia}

\affil[*]{d.bernhardt@uq.edu.au}

\begin{abstract}
  Fluctuation theorems (FTs) quantify the thermodynamic reversibility of a
  system, and for deterministic systems they are defined in terms of the dissipation function.
  However, in a nonequilibrium steady state of deterministic dynamics, the
  phase space distribution is unknown, making the dissipation function difficult
  to evaluate without extra information.
  As such, steady state FTs for deterministic systems to date have
  required either that the trajectory segment of interest is relatively long, or
  that information is available about the entire trajectory surrounding that segment.
  In this work, it is shown that a simple machine learning model trained to
  predict whether a given steady state trajectory segment is being played
  forward or backward in time calculates a function which satisfies an FT and
  relies solely on information within the segment of interest.
  The FT is satisfied even for very short trajectory segments where the approximate
  relation derived from theory breaks down, for systems far from
  equilibrium, and for various nonequilibrium dynamics.
  It is further demonstrated that any function which is a well-calibrated predictor of
  time's arrow must satisfy a fluctuation theorem, and that a local FT can be derived
  which depends only on local dissipation and its correlations with the surrounding
  non-local dissipation.

\end{abstract}

\begin{document}

\flushbottom \maketitle

\thispagestyle{empty}

\section{Introduction}

Nonequilibrium processes are ubiquitous in nature and everyday life, yet our
understanding of them from a thermodynamics perspective is limited.
One of the few thermodynamic relations that is valid arbitrarily far from
equilibrium is the fluctuation theorem (FT) that quantifies the probability of
observing, over some period of time, events that violate the
2\textsuperscript{nd} law of thermodynamics \cite{Evans2016}.
The FT of Evans and Searles \cite{Evans1994,Evans2002}, which is applicable to deterministic dynamics,
has been used to derive a number of useful relations.
This includes the Green-Kubo relations for transport properties, and the dissipation theorem, which gives
the average value of a phase function arbitrarily near or far from equilibrium and can achieve greater
accuracy over shorter time periods or under smaller driving forces in comparison to direct averaging techniques \cite{Evans2016, Evans2008a, Evans2008b, Bernardi2012, Maffioli2022}.
However, the FT for deterministic dynamics can only be exactly evaluated under the condition
that the distribution function of the initial ensemble from which the nonequilibrium trajectories
begin is a known (typically equilibrium) one.
Often we are interested in nonequilibrium steady state systems which are either approximately characterised by a continuously
changing phase distribution function after the phase variables have  reached their steady state values to within a desired tolerance, or by a steady state measure that might not be a function.
For these systems, if only time-local information is known, the FT can only be
evaluated as an approximate relation that is asymptotically valid as the duration of the measurement increases \cite{Evans2016}.

The broad umbrella of nonequilibrium steady state systems encompasses many important processes across various fields, including fluid flow, heat transfer, semiconductor operation, ion transport, biological phenomena,
and climate modelling \cite{Yesilata2006,Paul2010,Palacios2019,Bardeen2014,Balsara2015,Vogel2012,Montgomery2017}.
For many systems to function, it is necessary that they are maintained in a state other than the equilibrium one, requiring application of a constant driving force or flux and leading to a nonequilibrium steady state.
Hence, the discovery of improved theory for treating nonequilibrium steady states could have a wide impact.

With the growing capabilities of machine learning and artificial intelligence,
it has become apparent that such models can prove an effective tool for gaining new
and interesting scientific understanding \cite{Carleo2019,Krenn2022}.
Notably, it was recently demonstrated by Sief \textit{et al.} that a machine learning model is capable of
rediscovering the Crooks FT for transient stochastic dynamics \cite{Seif2021}.
In this work, we apply simple machine learning techniques to non-equilibrium steady states.
Our model discovers a steady state fluctuation theorem for deterministic dynamics that remains valid
even for very short measurements and is more accurate than previously known relationships.

\section{Background}

\subsection{Steady-state fluctuation theorems}
The fluctuation theorem quantifies the reversibility of a system
as \cite{Evans2002}
\begin{equation}
  \ln\frac{p(\Omega_{0,t}(\Gamma;0)=At)}{p(\Omega_{0,t}(\Gamma;0)=-At)} \equiv \ln\frac{p(\bar{\Omega}_{0,t}(\Gamma;0)=A)}{p(\bar{\Omega}_{0,t}(\Gamma;0)=-A)}=At,
  \label{eqn:FT}
\end{equation}
where $\Omega_{0,t}(\Gamma;0)$ is the dissipation function of the trajectory of the phase space vector,
$\Gamma$, integrated between time 0 and time $t$, defined with respect to the
phase space distribution at time 0, and $p(\Omega_{0,t}(\Gamma;0)=At)$ is the
probability that a trajectory has the particular dissipation value $At$.
The symbol $\bar{\Omega}$ denotes the time averaged value over the indicated segment.

In general, the time integral of the dissipation function of a trajectory segment from time $t_1$ to
time $t_2$ is given by
\begin{equation}
  \Omega_{t_1,t_2}(\Gamma; t_d) \equiv \ln\frac{f(\Gamma_{t_1};
  t_d)}{f(M^T\Gamma_{t_2}; t_d)} - \int_{t_1}^{t_2}\Lambda(\Gamma_s)ds,
  \label{eqn:DFunc}
\end{equation}
where $f(\Gamma_t; t_d)$ is the phase space probability density of the system at time $t_d$, evaluated at the point $\Gamma_t$.
$\Gamma_t$ is a point in phase space at time $t$, which is independent of $t_d$.
$M^T$ is a time-reversal mapping,%
\footnote[2]{More generally, this is the mapping required so that when the system is subject to the same driving force, the dynamics will generate a trajectory with an integrated dissipation of the same magnitude but opposite sign. The mapping is often $\{ \textbf{q},\textbf{p} \}  \rightarrow \{  \textbf{q},-\textbf{p} \} $, but exceptions include shear flow \cite{Evans1996}, considered in Section \ref{shearflow}, and when a magnetic field is considered\cite{Coretti2021}.  } 
and
\begin{equation}
  \Lambda(\Gamma) \equiv \frac{\partial}{\partial\Gamma}\cdot\dot{\Gamma}
\end{equation}
is the phase space expansion factor \cite{Evans2002}.
For the dissipation function to satisfy the fluctuation theorem in the form of Equation \ref{eqn:FT}, it must be
integrated from the same time as the distribution relative to which it is
defined; that is, $t_d=t_1$.
Furthermore, in order to evaluate the dissipation function, the phase space
distribution function at $t_d$ must be well defined, which is not the case for
trajectory segments beginning from a nonequilibrium steady state.

There are two common approaches towards obtaining the steady state FT (SSFT) from consideration of transients (trajectories beginning from $t=0$).
Firstly, consider the case where the distribution function is known at $t=0$ (\textit{e.g.} it is an
equilibrium distribution) and the system is evolving towards a steady state.
Noting that \cite{Evans2016}
\begin{equation}
  \Omega_{t,t+\tau}(\Gamma;t) = \Omega_{0,2t+\tau}(\Gamma;0),
\end{equation}
and hence,
\begin{equation}
  \tau\bar{\Omega}_{t,t+\tau}(\Gamma;t) =
  (2t+\tau)\bar{\Omega}_{0,2t+\tau}(\Gamma;0),
  \label{eqn:split trajectory}
\end{equation}
one can use knowledge of the surrounding periods of the trajectory to define an exact SSFT as \cite{Evans2016}
\begin{eqnarray}
  \ln\frac{p\left(\bar{\Omega}_{t,t+\tau}(\Gamma;t)=A\right)}
  {p\left(\bar{\Omega}_{t,t+\tau}(\Gamma; t)=-A\right)} 
  &=&
  \ln\frac{p\left(\bar{\Omega}_{0,2t+\tau}(\Gamma;0)=\frac{A\tau}{2t+\tau}\right)}
  {p\left(\bar{\Omega}_{0,2t+\tau}(\Gamma;0)=-\frac{A\tau}{2t+\tau}\right)} \nonumber\\
  &=& A\tau.
\end{eqnarray}
Although this is an exact expression, it is generally not of practical use.
This is because the dissipation function on the left hand side is defined with respect
to the time-evolved distribution at time $t$, and the dissipation function on the right
hand side requires knowledge of the trajectory outside the period of interest.
If only time-local information is available and/or $t\rightarrow\infty$, this is problematic.

Alternatively, noting that
\begin{equation}
  \Omega_{t,t+\tau}(\Gamma;0)=\Omega_{0,\tau}(\Gamma;0) +
  \Omega_{\tau,\tau+t}(\Gamma;0)-\Omega_{0,t}(\Gamma;0),
\end{equation}
if we can assume that correlations decay sufficiently quickly (the system is T-mixing \cite{Evans2016}) and select $t$ to be much greater than $\tau_M$, the Maxwell time of the system which characterises time correlations, then one can make the approximation that \cite{Evans2016}
\begin{equation}
  \bar{\Omega}_{t,t+\tau}(\Gamma;0) =
  \bar{\Omega}_{0,\tau}(\Gamma;0) + \mathcal{O}\left(\frac{\tau_M}{\tau}\right).
\end{equation}
Hence, the approximate SSFT can be derived as \cite{Evans2016}
\begin{equation}
  \lim_{\tau\rightarrow\infty}\frac{1}{\tau}\ln\frac
  {p(\Omega_{t,t+\tau}(\Gamma;0)=A\tau)}
  {p(\Omega_{t,t+\tau}(\Gamma;0)=-A\tau)} = A,
  \label{eqn:ss_dissipation_function_approximation}
\end{equation}
provided the standard deviation of the distribution of values, $A$, shrinks less quickly than $1/\tau$.
Then, the steady state dissipation function defined with respect to the
initial, known distribution satisfies the fluctuation theorem in the limit
that the trajectory segment of interest is much longer than the correlation
time.
However, for small $\tau$ this limit does not apply.

A similar problem is encountered when deriving the SSFT from the functional FT \cite{Ayton2000,Searles2013,Evans2016},
\begin{equation}
    \frac{p(F(\Gamma_{0\rightarrow2t+\tau})=A)}{p(F(\Gamma_{0\rightarrow2t+\tau})=-A)}
        = \left<e^{-\Omega_{0,2t+\tau}(\Gamma;0)}\right>_{F(\Gamma_{0\rightarrow2t+\tau})=-A},
    \label{eqn:functional_ft}
\end{equation}
where $\Gamma_{0\rightarrow2t+\tau}$ denotes the trajectory segment from time $0$ to time $2t+\tau$, $F$ is a path function of that segment that is odd with respect to time-reversal, and $\left<\dots\right>_{\text{cond.}}$ denotes an ensemble average over trajectory segments that satisfy the given condition.
Substituting $F(\Gamma_{0\rightarrow2t+\tau})=\Omega_{t,t+\tau}(\Gamma;0)$ gives \cite{Searles2013}
\begin{equation}
    \frac{p(\Omega_{t,t+\tau}(\Gamma;0)=A\tau)}{p(\Omega_{t,t+\tau}(\Gamma;0)=-A\tau)} =
        e^{A\tau}\left<e^{-\Omega_{0,t}(\Gamma;0)}e^{-\Omega_{t+\tau,2t+\tau}(\Gamma;0)}\right>_{\Omega_{t,t+\tau}(\Gamma;0)=-A\tau}.
\end{equation}
For a T-mixing system with $\tau$ much greater than the Maxwell time,
correlations between $\Omega_{0,t}(\Gamma;0)$, $\Omega_{t,t+\tau}(\Gamma;0)$ and $\Omega_{t+\tau,2t+\tau}(\Gamma;0)$ can be assumed to be independent of $\tau$, in which case
\begin{align}
    \left<e^{-\Omega_{0,t}(\Gamma;0)}e^{-\Omega_{t+\tau,2t+\tau}(\Gamma;0)}\right>_{\Omega_{t,t+\tau}(\Gamma;0)=-A\tau}
        =& \left<e^{-\Omega_{0,t}(\Gamma;0)}\right>_{\Omega_{t,t+\tau}(\Gamma;0)=-A\tau}\times\hfill\nonumber\\
        &\left<e^{-\Omega_{t+\tau,2t+\tau}(\Gamma;0)}\right>_{\Omega_{t,t+\tau}(\Gamma;0)=-A\tau}\nonumber\\
        =& C(t),
\end{align}
again resulting in Equation \ref{eqn:ss_dissipation_function_approximation}.

The problems encountered here are not unique to steady states, but are more
generally reflective of difficulties in treating a dissipation function which
is missing information.
The initial distribution is unknown for trajectories beginning from a steady state,
as is their time evolution from the known ($t=0$) distribution in the case of
time-local measurements.
Similar difficulties are found when deriving the space-local FT for
the dissipation function calculated over a local volume of a larger system,
in which case the missing information is the dissipation function of the
remaining volume of the system.
However, although there has been some progress towards space-local fluctuation theorems
through both theoretical and experimental investigations
\cite{Talaei2012,Michel2013,Shang2005,Ayton2001,Feitosa2004,Gallavotti1999},
there is currently no evaluable time-local SSFT that remains accurate for
arbitrarily short trajectories.

It has been shown that if a local dissipation function is defined as
$\Omega_{0,t}^L(\Gamma;0)=\Omega_{0,t}(\Gamma;0)-\Omega_{0,t}^{L'}(\Gamma;0)$,
where $\Omega_{0,t}^{L'}(\Gamma;0)$ is the remaining, non-local component of
the global dissipation function, then for some systems an approximate space-local
FT can be obtained using a linear scaling factor $\alpha$ \cite{Michel2013,Talaei2012}.
This scaling factor accounts for first-order correlations between the local and non-local
components such that
\begin{equation}
  \Omega_{0,t}^{L'}(\Gamma;0) = \alpha\Omega_{0,t}^L(\Gamma;0) + \xi,
  \label{eqn:local_ft_splitting}
\end{equation}
where $\xi$ represents the uncorrelated components of
$\Omega_{0,t}^{L'}(\Gamma;0)$.
This gives the space-local FT \cite{Michel2013,Talaei2012},
\begin{equation}
  \ln\frac{p(\Omega_{0,t}^L(\Gamma;0)=At)}{p(\Omega_{0,t}^L(\Gamma;0)=-At)}
    = (1+\alpha)At.
    \label{eqn:space_local_FT}
\end{equation}
If spatial correlations are then assumed to be exponentially decaying in nature,
a value of $\alpha$ can be calculated as \cite{Michel2013,Talaei2012}
\begin{equation}
  \alpha \approx
    \frac{l_0\left(1-e^{-l/l_0}\right)}{l-l_0\left(1-e^{-l/l_0}\right)},
    \label{eqn:space_local_correlations}
\end{equation}
where $l_0$ is the correlation length, and $l$ is the size of the local area of
interest for which $\Omega_{0,t}^L(\Gamma;0)$ is defined.
Although this formalism may, in principle, extend to time-locality, time
correlations in $\Omega(\Gamma;0)$ may not be
well represented by an exponential decay, making $\alpha$ difficult to
calculate.
Hence, an alternative approach is required.

\subsection{Machine learning fluctuation theorems}
One interesting property of the fluctuation theorem is that in theory it can be
used to predict whether a movie of a nonequilibrium physical process is being
played forward or backward.
Indeed, it was recently shown by Seif \textit{et al.} \cite{Seif2021} that a neural
network recovers a form of the Crooks FT \cite{Crooks1999} when given an equal mix of
forward and reverse, stochastic, transient trajectories as input and trained to
predict the direction of time's arrow.
In the present work, similar machine learning methods are employed to
investigate the possibility of an accurate time-local SSFT for deterministic
dynamics.

Following Crooks and Jarzynski \cite{Crooks1999, Jarzynski2011}, for a trajectory of
length $\tau$ sampled from an equal mix of forward-process
and time-reversed reverse-process trajectories, the probability
that it was generated by the forward process is given by
\begin{equation}
  p_+(\Gamma_{0 \rightarrow \tau}) = \frac{1}{1+e^{-\beta(W_{0,\tau}+\Delta F)}},
  \label{eqn:timearrow}
\end{equation}
where $W_{0,\tau}$ is the work along the trajectory from time 0 to time $\tau$,
$\beta^{-1}=k_BT$ is the inverse temperature, and $\Delta F$ is the free
energy difference between the initial and final configuration of the process.
See Section 1 of the supporting information for the derivation of this expression
from the Crooks FT.

It is worth clarifying here that the forward and reverse processes considered when
discussed with regard to Equation \ref{eqn:FT} are not necessarily
synonymous with those considered when discussing the
FT of Equation \ref{eqn:timearrow} \cite{Evans2016}.
Consider trajectories that are sampled from an initial equilibrium distribution characterised by $\lambda_A$, with $\lambda(t)$ being a parameter describing a time-dependent process driving a change in the system, with $\lambda(\tau)=\lambda_B$.
A set of trajectories, $\Gamma_{0 \rightarrow \tau}$, progress from the initial equilibrium ensemble to some final
(nonequilibrium) ensemble of states at $\tau$.
There will be a corresponding set of conjugate trajectories, $\Gamma^*_{0 \rightarrow \tau}$, that are related by time-reversal symmetry.
If the protocol for variation of $\lambda(t)$ is not time-symmetric over the period $\tau$, then the conjugate trajectories will be generated with the evolution of $\lambda(t)$ reversed (varying from $\lambda_B$ to $\lambda_A$).
The FT given by Evans and Searles, which is stated in Equation \ref{eqn:FT}, quantifies the probability ratio of observing,
from the {\textit{same}} initial phase space distribution, a trajectory, $\Gamma_{0 \rightarrow \tau}$, and its
corresponding conjugate trajectory, $\Gamma^{*}_{0 \rightarrow \tau}$.
\footnote[3]{Note, if the protocol for variation of $\lambda$ is not time-symmetric, then the set of trajectories considered in evaluation of the probabilities in Equation \ref{eqn:FT} will include those generated with both the forward and reverse evolution of $\lambda$ (i.e. from $\lambda_A$ to $\lambda_B$ and from $\lambda_B$ to $\lambda_A$). }
In contrast, the Crooks FT compares the probabilities of trajectories $\Gamma_{0 \rightarrow \tau}$ sampled from an initial equilibrium distribution characterised by $\lambda_A$ and driven by $\lambda(t)$ changing to $\lambda_B$, with the probabilities of conjugate trajectories $\Gamma^*_{0 \rightarrow \tau}$ sampled from the equilibrium distribution characterised by $\lambda_B$ and $\lambda(t)$ changing to $\lambda_A$ with the reverse protocol.  

As we are interested in deterministic systems that are subject to a constant driving force, not the
transition between two equilibrium states, in this work we deal exclusively with
systems for which there is no free energy difference and
the time-local version of the Evans-Searles FT is an approximate relation (Equation
\ref{eqn:ss_dissipation_function_approximation}).
The systems we consider here are generated by time-reversible and deterministic dynamics,
which differs from the equations of motion of reference \cite{Seif2021} that are not
time-reversible.
Therefore it might be supposed that it is impossible to determine if the trajectory is more
likely to have been sampled from a forward or reverse simulation.
However, it has been observed that for steady states there is a time asymmetry in the fluctuations \cite{Paneni2006,Paneni2008}, and a steady state FT is followed \cite{Evans1994,Wang2005,Searles2013}.

To expand on this, we note that for this type of dynamics (deterministic and reversible), any trajectory that is a solution to the equations of motion will be possible in either the forward or reverse process.
So the question is not whether the trajectory was played forwards or backwards (as both are possible), but the likelihood that it was generated naturally, or after being played in reverse.
See Figure S1 in the supporting information for an illustration.
This is different to irreversible dynamics where a trajectory that occurs in the forward direction could be impossible in the set of trajectories played in reverse, and vice versa.  

Therefore, considering Equations \ref{eqn:split trajectory} and \ref{eqn:ss_dissipation_function_approximation} and Section 1 of the supporting information, the
probability that a trajectory is a forward one, when drawn randomly
from an even mix of forward and time-reversed trajectories, is given by
\begin{eqnarray}
  p_+(\Gamma_{t\rightarrow t+\tau})
      &=& \frac{1}{1+e^{-\Omega_{t,t+\tau}(\Gamma;t)}}\nonumber\\
      &=& \frac{1}{1+e^{-\Omega_{0,2t+\tau}(\Gamma;0)}}\nonumber\\
      &\approx & \frac{1}{1+e^{-\Omega_{t,t+\tau}(\Gamma;0)}},
      \label{eqn:approxESprediction}
\end{eqnarray}
where the final result requires $\tau \gg \tau_M$, except when $t=0$ when it becomes an equality. 

\section{Machine learning model}
\subsection{Structure}

As our aim is to determine whether a machine learning model can find a new,
more accurate form of the time-local SSFT for deterministic systems, we begin
by constructing a model which predicts whether a given deterministic
trajectory segment is being played forwards or backwards in time \cite{Seif2021}.
This binary classification model should output a value between 0 and 1, with
values closer to 1 being forward predictions, while those closer to 0 are
reverse predictions.
Such bounding was achieved, as is common practice for binary classifiers, by applying
the nonlinear sigmoid function,
\begin{equation}
    \sigma(x)=\frac{1}{1+e^{-x}}, \label{eqn:sigmoid}
\end{equation}
to the scalar output of the final neural network layer.
Note that this is an arbitrary choice in that any sufficiently complex combination
of nonlinear functions could theoretically produce an equivalent result, although
a more complex neural network would likely be required in such cases to obtain
equivalent accuracy.

Whereas initial models used raw trajectories (time-dependent particle positions
and momenta) as input, a more complex model was required for more complicated
dynamics, as was observed previously in reference \cite{Seif2021}.
It was found that by instead using samples of the instantaneous dissipation
function, $\Omega(\Gamma;0)$, as input, which is defined by
\begin{equation}
    \Omega_{t,t+\tau}(\Gamma;0) = \int_t^{t+\tau}\Omega(\Gamma_s;0)ds,
    \label{eqn:instantaneous_dissipation}
\end{equation}
good results could be obtained even with a simple single-layer logistic
regression model for all dynamics tested.

With this single-layer structure, the probability assigned by the
model that a trajectory is a forward one can be written as
\begin{equation}
  p_+^{NN}(\Gamma) = \frac{1}{1+e^{-B_{t,t+\tau}(\Gamma)}},
\end{equation}
where, for input generated by molecular dynamics simulations with an
integration timestep of $\delta t$,
\begin{equation}
  B_{t,t+\tau}(\Gamma) = \sum_{i=1}^Mw_i\Omega\left(\Gamma_{t+i\delta t};0\right),
\end{equation}
where $M=\tau/\delta t$.
Note that although initial tests included a constant bias in this equation as an
extra free parameter, it was found that the bias was approximately 0 in all cases,
and averaged to 0 when models were re-trained multiple times.
This was as expected for a process without a free energy change when the sigmoid
function is used for nonlinearity with this simple model, and hence the bias
was not included in the model for any results presented here, leaving only
the weights as free parameters.

\begin{figure*}[t]
  \center\includegraphics[width=\textwidth]{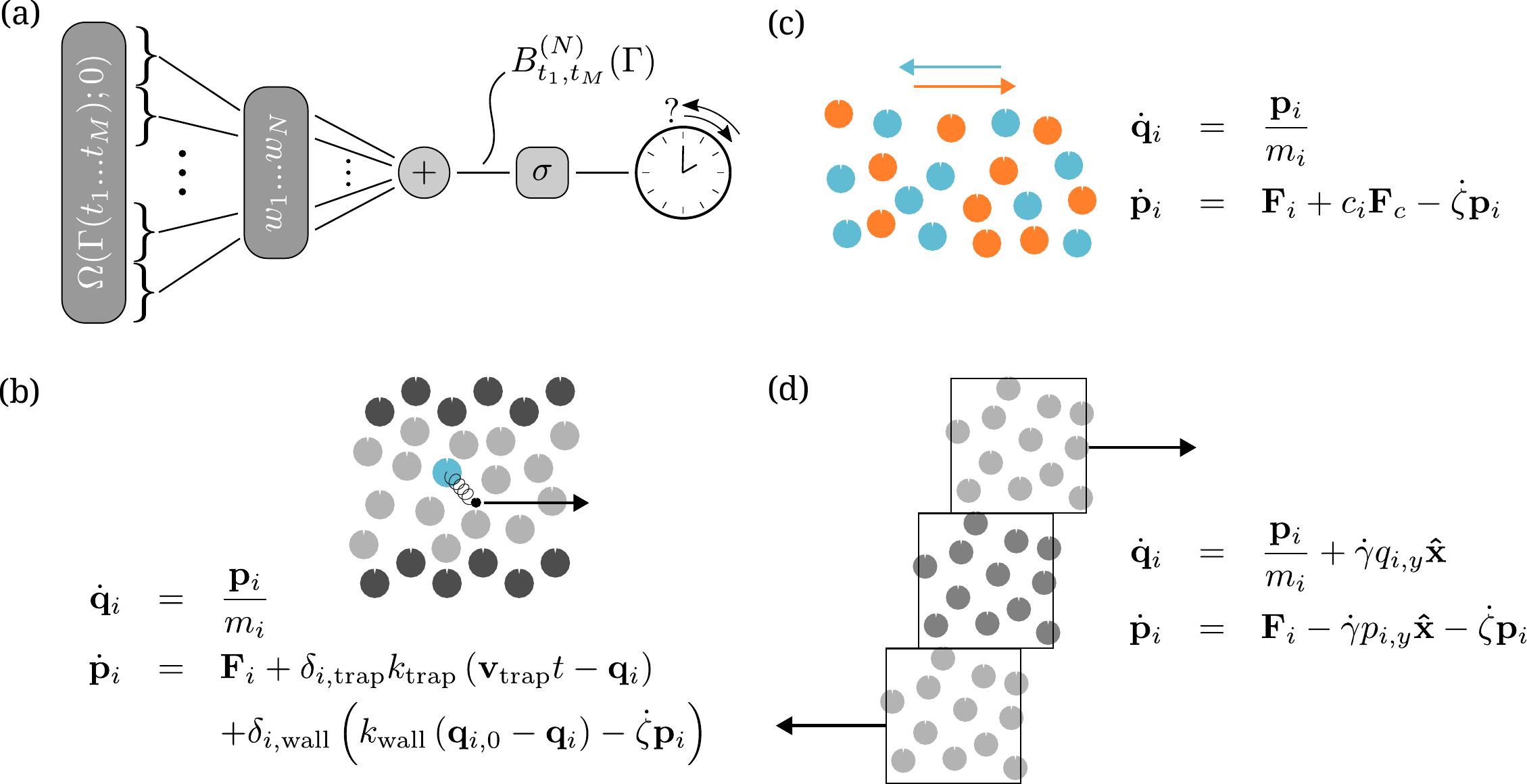}
  \caption{%
    Graphical representation of (a) the logistic regression model, (b) the optical tweezers system in which the wall (darker) particles are harmonically restrained to a lattice, (c) the color field system, and (d) the planar shear system illustrating Lees-Edwards boundary conditions applied to the periodic images above and below the unit cell.
    The instantaneous dissipation function (with respect to the known initial
    distribution) was used as input in (a), and the sigmoid function was used for
    nonlinearity to allow simple comparison with the fluctuation theorem.
    $\mathbf{q}_i$ and $\mathbf{p}_i$ represent the position and momentum, respectively, of particle $i$, and $\delta$ is the Kronecker delta function.
    $\dot\zeta$ is the thermostat coefficient governed by $\ddot{\zeta} = (2K_{\text{tp}}/(dN_{\text{tp}} k_\text{B}T) - 1)/\tau_{\text{damp}}^2$ where $\tau_{\text{damp}}$ is the time constant of the thermostat, $d$ is the dimensionality of the system, $N_{\text{tp}}$ is the number of thermostatted particles, equal to the number of wall particles in (a) or the total number of particles in (b) and (c), and $K_{\text{tp}}$ is the kinetic energy of those particles.
    In (b), $\mathbf{q}_{i,0}$ denotes the initial lattice position of wall particle $i$.
    In (c), $q_{i,y}$ and $p_{i,y}$ denote the $y$ component of the position and momentum of particle $i$, and $\mathbf{\hat{x}}$ represents the $x$ unit vector.
    Note that the optical tweezers system was simulated in 2D, while the color field and shear flow systems were simulated in 3D.}
  \label{fig:graphical_models}
\end{figure*}

To reduce the number of free parameters to $N$ (and thereby reduce the amount of
data required for good convergence), each consecutive set of $M'=M/N$ input values
was summed before applying the weights, giving
\begin{equation}
  B^{(N)}_{t,t+\tau}(\Gamma)
  =\sum_{i=1}^{N}w_i\sum_{j=1}^{M'}\Omega\left(\Gamma_{t+[(i-1)M'+j]\delta t};0\right),
  \label{eqn:B_sums}
\end{equation}
and the associated probability
\begin{equation}
  p_+^{NN}(\Gamma) = \frac{1}{1+e^{-B^{(N)}_{t,t+\tau}(\Gamma)}}. \label{eqn:nn_probability}
\end{equation}
See Figure \ref{fig:graphical_models}a for a graphical representation of the full
model structure, and Figure S2 in the supporting information for confirmation
that such a model is capable of reproducing the results of Seif \textit{et al.}
\cite{Seif2021} for the transient optical tweezers experiment in the case of
deterministic dynamics.

Noting that
\begin{equation}
  \int_t^{t+M'\delta t}\Omega(\Gamma_s;0)ds
  = \delta t\sum_{j=1}^{M'}\Omega(\Gamma_{t+j\delta t};0)
  + \mathcal{O}\left(\delta t^2\right),
\end{equation}
it is clear from Equation \ref{eqn:instantaneous_dissipation} that for small $\delta t$, Equation \ref{eqn:B_sums} is equivalent to
\begin{equation}
    B^{(N)}_{t,t+\tau}(\Gamma)\approx\sum_{i=1}^N \frac{w_i}{\delta t}\Omega_{t_{i-1},t_i}(\Gamma;0),
    \label{eqn:nn_dissipation_integration_approx}
\end{equation}
where $t_i=t+iM'\delta t$.
Hence, the input to the model is, to a first order approximation, equivalent
to consecutive, equal-length segments of the time-local dissipation function.

\subsection{Training}
Each data set was split into a training set (typically of 12,000 trajectories), a
validation set (typically of 4,000 trajectories), and a test set (typically
84,000 trajectories).
In each data set, exactly half of the trajectories were randomly chosen to be
time-reversed ones.
As the dynamics are deterministic, time reversal could be achieved simply
by reversing the order of samples of $\Omega(\Gamma;0)$ and inverting the sign;
that is, for a particular phase point, $\Gamma_{t+s}$, at time $s$ after the beginning
of the trajectory segment $\Gamma_{t\rightarrow t+\tau}$,
$\Omega(\Gamma_{t+s};0) = -\Omega(\Gamma_{t+\tau-s}^{*};0)$.

The machine learning model was trained using the Adam optimiser
\cite{Kingma2015} to minimise the binary cross-entropy loss function
\cite{Seif2021,Goodfellow2016}.
Note that while initial results were obtained using the $L_2$ regularization
technique to prevent over-reliance on a single weight \cite{Goodfellow2016}
this was ultimately disabled to ensure that it did not affect the distribution
of the weights.
The lack of weight over-reliance prevention did not noticeably affect the
accuracy or rate of convergence for the systems tested here.

\section{Test systems}
Input data was generated using deterministic molecular dynamics simulations
of various nonequilibrium processes as outlined below.
In all cases, the equations of motion were integrated using the 4th order
Runge-Kutta method, and quantities are given in Lennard-Jones normalised units.
Unless otherwise stated, the integration timestep was $\delta t = 0.001$, and
temperature was controlled using a Nosé-Hoover thermostat with $k_BT=1.0$ and a
time constant of $100\delta t$ \cite{Hoover1985}.
Initial particle velocities were sampled from a Boltzmann distribution at the
same temperature as the thermostat, with any global momentum bias subtracted
before beginning the simulation.

\subsection{Optical tweezers}
Firstly, the 2D optical tweezers experiment was used, as described in detail by
Wang \textit{et al.} \cite{Wang2002} and represented graphically in Figure \ref{fig:graphical_models}b.
Only the wall particles were influenced by the thermostat, while the motion of
the fluid particles was according to Newton's laws.
A total of 44 wall particles were used, and they were harmonically restrained to
their initial locations in a hexagonal lattice of number density 1.0 with a
spring constant $k_{\text{wall}}=150$.
Fluid particles were initialised as part of the same lattice, and were then
randomly deleted to achieve the desired fluid number density ($\rho$).
The periodic boundaries parallel and perpendicular to the walls were
approximately of length 11.2 and 11.8 respectively.
Particle interactions were of the Weeks-Chandler-Anderson (WCA) type \cite{Weeks1971}.
One fluid particle close to the middle of the channel was chosen to be
influenced by a harmonic trap (of spring constant $k_\text{trap}$), which was
located equidistant from the walls.
The system was then equilibrated for 1 million timesteps before beginning the
nonequilibrium process, at which point the harmonic trap began moving parallel
to the walls with constant velocity $\mathbf{v}_{\text{trap}}$.
The instantaneous dissipation function in this experiment is given by
\begin{equation}
  \Omega(\Gamma;0) = -\beta\mathbf{F}_{\text{trap}}\cdot\mathbf{v}_{\text{trap}},
\end{equation}
where $\mathbf{F}_{\text{trap}}$ is the force on the trapped particle due to the
harmonic restraint \cite{Wang2002}.

\subsection{Color field}
Second was a color field experiment, shown pictorially in Figure \ref{fig:graphical_models}c,
in which half of the particles were assigned a color `charge', $c_i$, of -1 while the other half are assigned +1.
These color charges interact only with an external field, $\mathbf{F}_c$,
driving differently colored particles in opposite directions to each other
with a force of $c_i\mathbf{F}_c$ \cite{Evans2008a,Sarman1998}.
Particle interactions were again of the WCA type, with a total of $N_p=96$
particles in a 3D, periodic, cubic simulation box scaled to give a number density of $\rho=0.7$.
All particles were influenced by the Nosé-Hoover thermostat, and were
equilibrated for 1 million timesteps before applying the external field.
The instantaneous dissipation function of this system is given by the `color
current',
\begin{equation}
  \Omega(\Gamma;0) = \beta\sum_{i=1}^{N_p} c_i \mathbf{v}_i \cdot \mathbf{F}_c,
\end{equation}
where $\mathbf{v}_i$ is the velocity of particle $i$.

\subsection{Shear flow}\label{shearflow}
Finally, 3D planar shear flow was simulated using the SLLOD algorithm
\cite{Evans2008a} with Lees-Edwards boundary conditions \cite{Lees1972},
as illustrated in Figure \ref{fig:graphical_models}d.
A shear rate of $\dot\gamma = \frac{dv_x}{dy}=0.1$ was used, with particles within a cut-off radius of 2.6 interacting via the 12-6 Lennard-Jones potential ($\sigma_{LJ}=1.0$, $\epsilon=1.0$).
128 particles were initialised in a hexagonal close-packed (HCP) lattice at a number
density of 0.6, with periodic dimensions of $5.32\times9.22\times4.35$ in the
$x$, $y$, and $z$ dimensions respectively ($4\times4\times2$ unit cells), giving
a total of 128 particles.
They were then allowed to equilibrate in the absence of shear (to the NVT
ensemble) for 10,000 timesteps before beginning the nonequilibrium protocol.
Here, the instantaneous dissipation function is given by
\begin{equation}
  \Omega(\Gamma;0) = -\beta\dot\gamma V P_{xy}(\Gamma_s),
\end{equation}
where $V$ is the volume of the simulation box, and $P_{xy}$ is the $xy$
component of the pressure tensor \cite{Evans2010}.

\section{Results and discussion}
\begin{figure}[tbh]
  \center\includegraphics[width=0.7\textwidth]{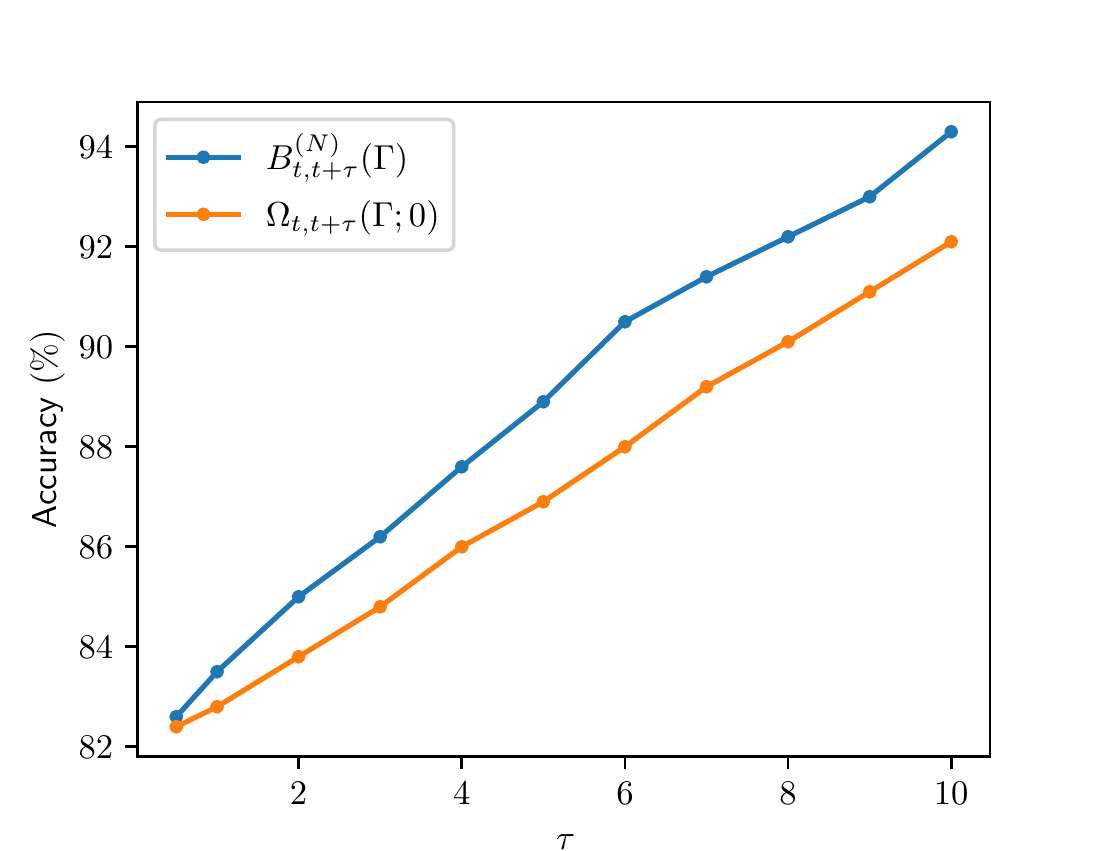}
  \caption{%
    Predictive accuracy of the logistic regression model as a function of $\tau$,
    compared against the accuracy when predictions were made from the expression
    $p_+(\Gamma)=\sigma(\Omega_{t,t+\tau}(\Gamma;0))$, as per the approximate
    SSFT.
    Input trajectories of length $\tau$ were generated from the 2D optical tweezers
    experiment with $\rho=0.4$, $v_\text{trap}=0.5$, $k_\text{trap}=1.0$, and
    $k_BT=1.0$.
    They were divided into segments such that $N=\tau/50\delta t$;
    that is, each input value to the network was the sum of the instantaneous
    dissipation function over 50 consecutive simulation timesteps.
  }
  \label{fig:ssft_accuracy}
\end{figure}

\begin{figure}
  \center\includegraphics[width=0.7\textwidth]{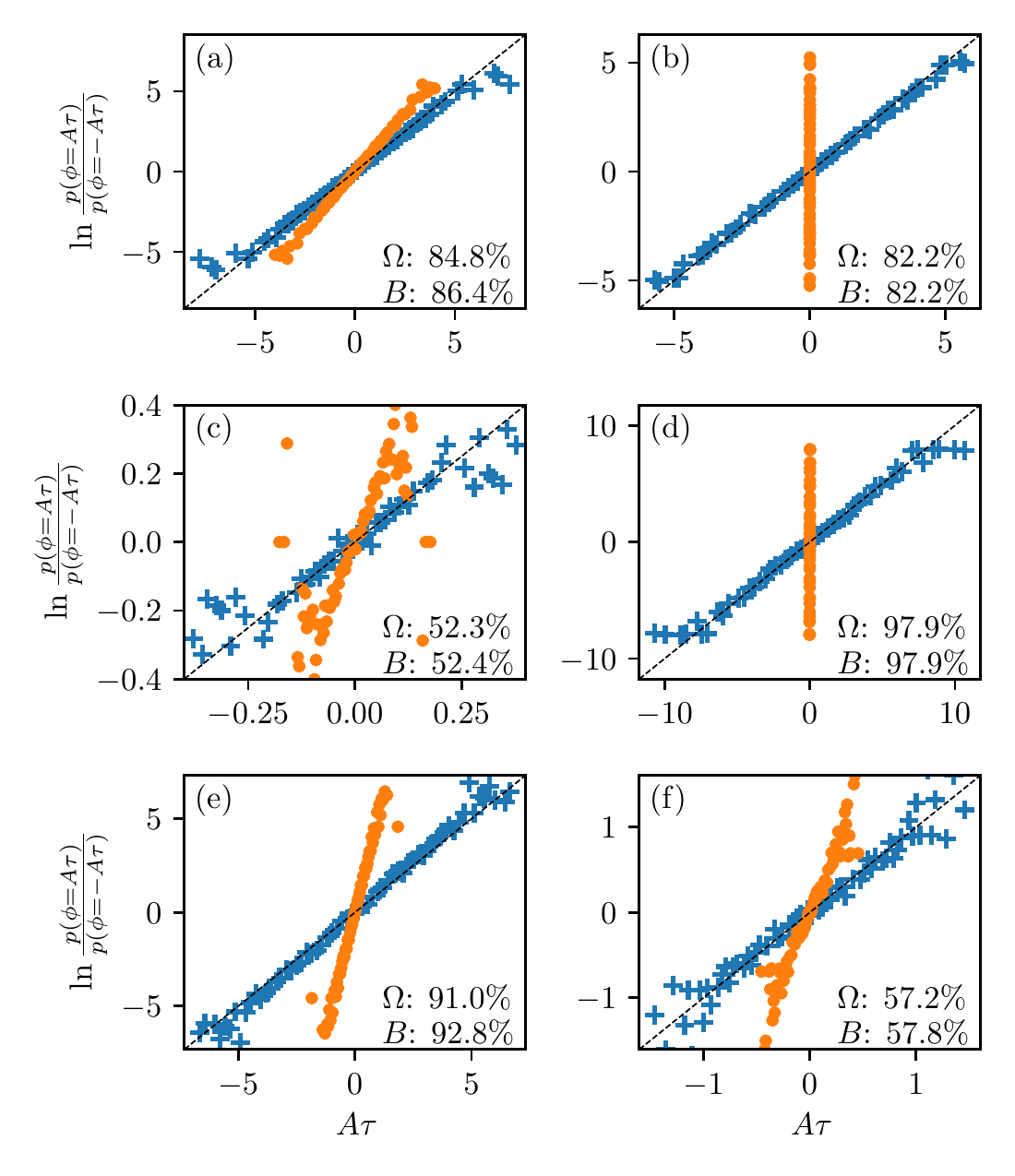}
  \caption{%
    Test of the FT comparing
    $\phi=\Omega_{t,t+\tau}(\Gamma;0)$ (orange dots), and the quantity
    calculated by the model, $\phi=B^{(N)}_{t,t+\tau}(\Gamma)$ (blue crosses), for (a)
    the optical tweezers experiment with $\rho=0.4$, $v_{\text{trap}}=0.5$
    $k_\text{trap}=1.0$, $\tau=3$ and $N=100$, and (b) the same experiment with a duration
    of $\tau=0.003$ and $N=3$.
    (c) and (d) compare the FT under the same experiment
    close to equilibrium ($v_{\text{trap}}=0.05$, $\tau=0.8$, $N=100$), and far from
    equilibrium ($v_{\text{trap}}=2.0$, $\tau=0.005$, $\delta t=0.0005$, $N=5$)
    respectively.
    (e) shows the same for the SLLOD equations of motion with
    $\dot\gamma=0.5$, $\tau=0.1$, $\rho=0.6$ and $N=100$, and (f) shows results
    for the color field experiment with $N=100$, $F_c=0.1$ and $\tau=0.1$ at a number density
    of $\rho=0.7$.
    Dashed lines are a guide to the eye, showing the expected relation.
    The predictive accuracies of $\Omega_{t,t+\tau}(\Gamma;0)$ and
    $B_{t,t+\tau}(\Gamma)$ are given in the lower right
    corner of each plot (note the magnitude is related to the distance from
    equilibrium and the duration).
  }
  \label{fig:FT_plots}
\end{figure}

We first discuss the case of applying the model to the optical tweezers
experiment.
It is immediately apparent from Figure \ref{fig:ssft_accuracy} that the model is
able to make predictions of time's arrow that are more accurate than what can be
achieved with $\Omega_{t,t+\tau}(\Gamma;0)$.
In addition, panels a and b of Figure \ref{fig:FT_plots} show that the value calculated by
the model appears to satisfy the FT,
\begin{equation}
  \ln\frac{p(B^{(N)}_{t,t+\tau}(\Gamma)=A\tau)}{p(B^{(N)}_{t,t+\tau}(\Gamma)=-A\tau)} = A\tau,
\end{equation}
even under conditions where the approximate FT for $\Omega_{t,t+\tau}(\Gamma;0)$
breaks down.
Since the steady state FT is only expected to be true in the limit of long $\tau$ \cite{Searles2013}, this difference is most apparent for very short durations.
For example, Figure \ref{fig:FT_plots}b shows that $B^{(N)}_{t,t+\tau}(\Gamma)$
satisfies the FT even for a trajectory of length 0.003 time units; much shorter
than the correlation time of $\Omega(\Gamma;0)$, and equivalent to only three
simulation timesteps ($\delta t=0.001$).
This suggests that the FT for $B^{(N)}_{t,t+\tau}(\Gamma)$ is valid for
arbitrarily short durations, which is interesting as it implies that the
neural network can learn the approximation that is made in obtaining the final line of Equation \ref{eqn:approxESprediction}.
The relation was also observed to be valid for systems closer to or further from
equilibrium (Figure \ref{fig:FT_plots}c-d), and for systems other than the
optical tweezers experiment (Figure \ref{fig:FT_plots}e-f).
Note that although using fewer than 3 simulation timesteps was found to result
in some error in the FT, the same dynamics and duration under a smaller integration
timestep ($\delta t=0.0005$) still produced a model that satisfied the FT.
Hence, the error can be attributed to the approximate integration of the
dissipation function in Equation \ref{eqn:nn_dissipation_integration_approx}.

Similar arguments to those presented in reference \cite{Petersen2013} indicate that for systems where $\tau$ is less than or similar to $\tau_M$, correlations with periods either side of the segment $\Gamma_{t \rightarrow t+\tau}$ might be expected to result in a FT of the form
\begin{align}
\hspace{-1.5em}\ln\frac{p(\Omega_{t,t+\tau}(\Gamma;0)=A\tau)}
  {p(\Omega_{t,t+\tau}(\Gamma;0)=-A\tau)} \approx A(\tau+c\tau_M)=A\tau(1+c\frac{\tau_M}{\tau}),
  \end{align}
where $c$ is a constant, at least for values of $A$ that are close to the mean.
This is equivalent to the introduction of a scaling factor, $(1+\alpha)$, and corresponds to the approach assumed and demonstrated in the space-local
FT (Equation \ref{eqn:space_local_FT}) \cite{Michel2013,Talaei2012}.
Considering the plots in Figure \ref{fig:FT_plots}, this would result in straight lines of slope $1+c\tau_M/\tau$ which would deviate greatly from the time-asymptotic result of $1$ when $\tau$ is short, as is evident in the figure.
However, using the same argument, the predictive accuracy of the approximate SSFT would only change slightly.
This can be seen by considering how $p_+ (\Gamma_{t\rightarrow t+\tau})$ would change with $c$ under this approximation.
Thus the predictive accuracy of $\Omega_{t,t+\tau}(\Gamma;0)$  is similar to that of $B_{t,t+\tau}(\Gamma)$ even when the FT are very different, as is evident in the results for small $\tau$ presented in Figures \ref{fig:ssft_accuracy}, \ref{fig:FT_plots}b and \ref{fig:FT_plots}d.

For systems very far from equilibrium where $p(\Omega_{t,t+\tau}(\Gamma;0)<0)$
is very small, a model that satisfies the FT is still theoretically possible
(see the discussion below), but is not practically feasible to train, or even
to plot the FT, due to the enormous amount of data that would be required in
order to include the rare but necessary 2\textsuperscript{nd} law-violating
trajectories.
Similarly, for systems very close to equilibrium, (Figure \ref{fig:FT_plots}c
for example), training of the model becomes more difficult due to the large
amount of overlap in the distribution of the time-local dissipation function
between the forward and reverse sets of trajectories.
In the latter case, a combination of larger data sets and a slower learning rate was
required to obtain good results.

\subsection{Limited information}
\begin{figure}[t]
  \center\includegraphics[width=0.7\textwidth]{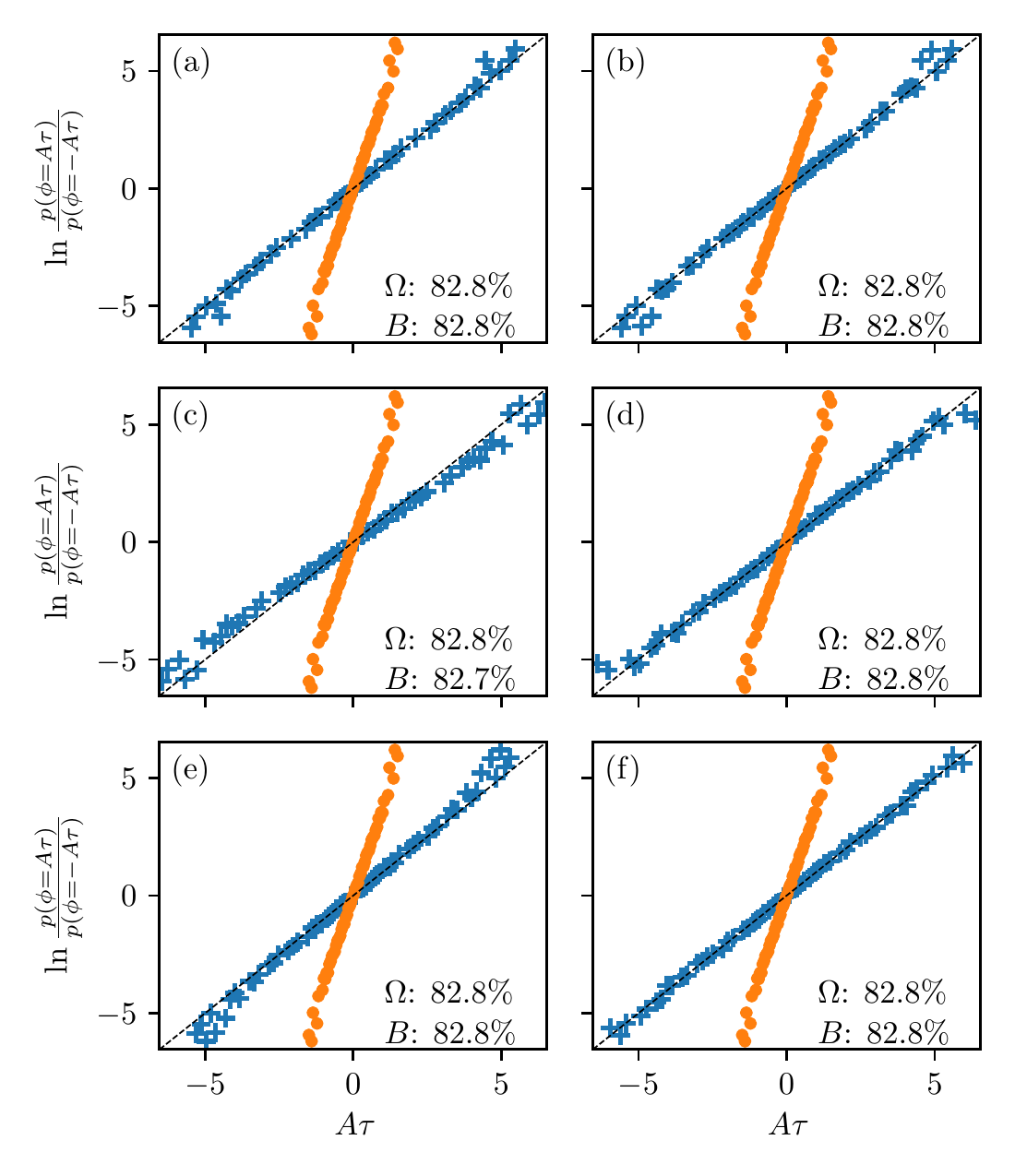}
  \caption{%
    Test of the FT comparing
    $\phi=\Omega_{t,t+\tau}(\Gamma;0)$ (orange dots), and the quantity
    calculated by the model, $\phi=B^{(N)}_{t,t+\tau}(\Gamma)$ (blue crosses) for the optical tweezers system with $\rho=0.4$, $\nu_{\text{trap}}=0.5$, $k_{\text{trap}}=1.0$, and $\tau=1$, where:
    (a) trajectories were randomly permuted each time they were loaded, with 20
    input segments per trajectory ($N=20$);
    (b) $N=1$;
    (c) weights randomly sampled from the distribution of weights obtained
    by re-training the model considered in (a) 100 times;
    (d) as for (c), but with weights re-sampled per-trajectory;
    (e) as for (c), but sampling weights from a Gaussian distribution with the
    same mean and variance; and
    (f) as for (e), but with weights re-sampled per-trajectory.
    All models were trained on the same data set, and tested on the same,
    separate data set.
  }
  \label{fig:FT_random_weights}
\end{figure}

To understand which features of the input are important, we now explore the effects of limited information
on the predictive accuracy, and on the accuracy with which the FT is satisfied.
First, models were trained under the condition that the order of the inputs were
randomly permuted for each trajectory, thereby removing information about causality.
We denote the value calculated by these
models as $B^{(N^*)}_{t,t+\tau}(\Gamma)$, and for models where the number of weights was less than the number of
timesteps ($M' > 1$), consecutive timesteps were first summed before random permutation.
Interestingly, this random permutation of the input had no obvious effect on the accuracy with
which the FT was satisfied, as can be seen in Figure \ref{fig:FT_random_weights}a.
However, it did limit the predictive accuracy to within margin of error of
$\Omega_{t,t+\tau}(\Gamma;0)$, suggesting that the higher predictive accuracy
of the initial model was achieved by taking advantage of the information about causality
of the input data.
This result also suggests that the predictive accuracy is only a
measure of the amount of information available, and optimal accuracy is not
required for the FT to be satisfied.

It was also observed that re-training the model multiple times on randomly
permuted data resulted in different weights each time, indicating that only the
distribution of the weights was important for the FT to be satisfied.
To confirm this, the network was re-trained 100 times on the same set of data
(with different random permutations each training epoch) to determine the cumulative
distribution function (CDF) of the weights.
A new set of weights was then sampled from that CDF and applied to the same test
data.
Figure \ref{fig:FT_random_weights}c shows that the FT was still approximately
satisfied, although not quite as accurately due to the possibility of an
`unlucky' set of weights that were not a good representation of the
distribution.
However, Figure \ref{fig:FT_random_weights}d shows that the effect of `unlucky'
sets of weights can be nullified by re-sampling the weights for each trajectory.
Interestingly, although replacing the measured CDF with a Gaussian distribution having
the same mean and variance and sampling a single set of weights does not accurately
satisfy the FT (Figure \ref{fig:FT_random_weights}e),
it was still satisfied if those Gaussian weights were re-sampled per trajectory
(Figure \ref{fig:FT_random_weights}f).

These results show that without information about causality, only the total
dissipation in the trajectory segment, $\Omega_{t,t+\tau}(\Gamma;0)$, is important
for the prediction, not the distribution of values.
This can be explained as follows.
Consider the value of $B^{(N)}_{t,t+\tau}(\Gamma_p)$ for any particular
trajectory given by $\Gamma_p$,
\begin{equation}
  B^{(N)}_{t,t+\tau}(\Gamma_p)
    = \sum_{i=1}^N \frac{w_i}{\delta t} \Omega_{t_{i-1},t_i}(\Gamma_p;0).
\end{equation}
Since the weights, $w_i$, are re-sampled periodically and are uncorrelated, the
average value for a given $\Gamma_p$ is
\begin{eqnarray}
  \left<B^{(N)}_{t,t+\tau}(\Gamma_p)\right>
    &=& \frac{\left<w\right>}{\delta t} \sum_{i=1}^N \Omega_{t_{i-1},t_i}(\Gamma_p;0),\nonumber\\
    &\approx& \frac{\left<w\right>}{\delta t}\Omega_{t,t+\tau}(\Gamma_p;0),
\end{eqnarray}
where the final line is an approximate relation following from Equation \ref{eqn:nn_dissipation_integration_approx}.
Hence, it is evident that sampling random weights with a mean value of $\left<w\right>$ is
equivalent to a single weight equal to $\left<w\right>$ in the limit of a large number of
samples.

Similarly, for a fixed set of weights, the average value of
$B^{(N)}_{t,t+\tau}(\Gamma)$ for the set of trajectories, $S$, with
$\Omega_{t,t+\tau}(\Gamma;0)=A\tau$ is given by
\begin{eqnarray}
  \left<B^{(N)}_{t,t+\tau}(\Gamma)\right>_{S}
    &=& \left<\sum_{i=1}^N \frac{w_i}{\delta t}\Omega_{t_{i-1},t_i}(\Gamma;0)\right>_{S},\nonumber \\
    &=& \sum_{i=1}^N \frac{w_i}{\delta t}\left<\Omega_{t_{i-1},t_i}(\Gamma;0)\right>_{S}.
\end{eqnarray}
If the input is randomly permuted, then at steady state
\begin{eqnarray}
  \left<B^{(N^*)}_{t,t+\tau}(\Gamma)\right>_{S}
    &=&
    \left<\Omega_{t,t+\tau/N}(\Gamma;0)\right>_{S}\sum_{i=1}^N \frac{w_i}{\delta t}
    \label{eqn:sep_from_rand_perm}\nonumber\\
    &=& \frac{A\tau}{N}\sum_{i=1}^N \frac{w_i}{\delta t},\nonumber\\
    &=& \frac{\left<w\right>}{\delta t}A\tau,
\end{eqnarray}
and hence this too is equivalent to a single weight in the limit of a large
number of samples.
Note, however, that in the case of fixed time ordering, correlations between the
segments of the dissipation function and the weights mean that the two cannot be
separated (i.e. Equation \ref{eqn:sep_from_rand_perm} would not necessarily be
correct), and therefore there is no equivalence to a single weight.
In this case, though, the predictive accuracy can exceed that of
$\Omega_{t,t+\tau}(\Gamma;0)$ if the weights bias forward trajectories such that
more have a positive value of $B^{(N)}_{t,t+\tau}(\Gamma)$ for a given value of
$\Omega_{t,t+\tau}(\Gamma;0)$.
That is, for trajectory segments in which the direction of time's arrow is
ambiguous as given by $\Omega_{t,t+\tau}(\Gamma;0)$, a predictor that biases
towards higher values for forward trajectories, or lower values for reverse
trajectories, will have a higher predictive accuracy.

\begin{figure}
    \centering
    \includegraphics[width=0.7\textwidth]{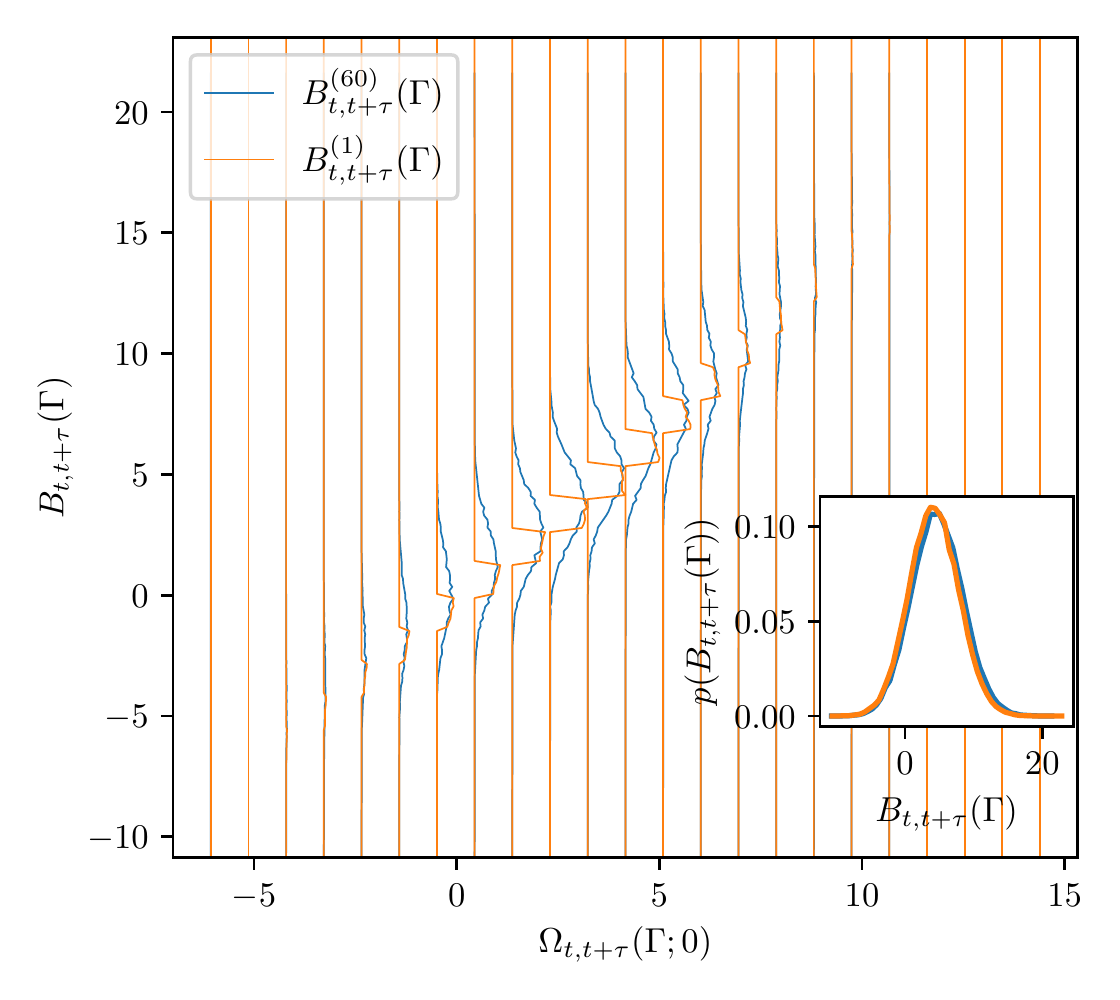}
    \caption{%
    Distribution of $B^{(N)}_{t,t+\tau}(\Gamma)$ given a particular value range of
    $\Omega_{t,t+\tau}(\Gamma;0)$ for the optical tweezers simulation with $\rho=0.4$,
    $\tau=6.0$ (see Figure \ref{fig:ssft_accuracy} for a comparison of predictive
    accuracies).
    Histograms in the main plot are normalised by the peak value for that
    data set, and are plotted as an offset to the $x$ value.
    The inset shows the overall distribution of $B_{t,t+\tau}(\Gamma)$ in
    each case (\textit{i.e.} the sum over all $\Omega_{t,t+\tau}(\Gamma;0)$ values).
    Note that $p(B^{(60)}_{t,t+\tau}(\Gamma)<0)$ is
    slightly lower than $p(B^{(1)}_{t,t+\tau}(\Gamma)<0)$,
    resulting in the slightly higher predictive accuracy of the former.
    Note also that reversing both axes gives the distribution for reverse
    trajectories, as does reversing the $x$ axis of the inset.
    }
    \label{fig:dissipation_hist}
\end{figure}

\begin{figure}
    \centering
    \includegraphics[width=0.7\textwidth]{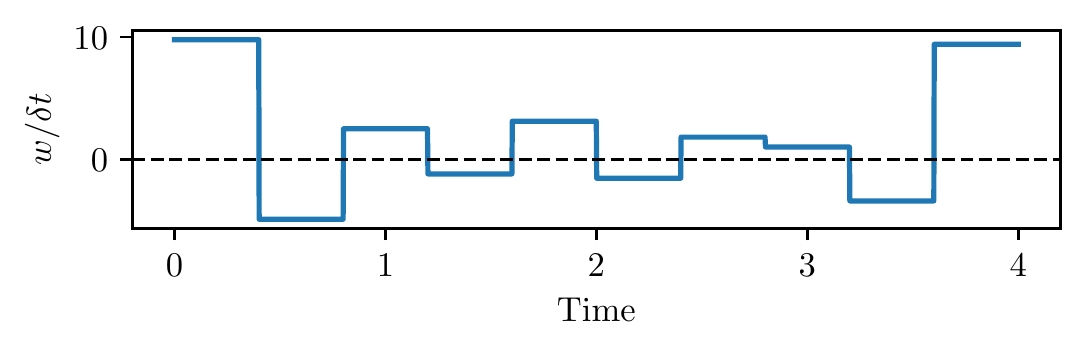}
    \caption{%
    Network weights as a function of the time segment they apply to
    for a model trained on the optical tweezers dynamics with $\rho=0.4$ and $\tau=4.0$
    (predictive accuracy shown in Figure \ref{fig:ssft_accuracy}).
    }
    \label{fig:nn_weght_dist}
\end{figure}

This phenomenon is demonstrated in Figure \ref{fig:dissipation_hist}, which
compares the distribution of $B^{(N)}_{t,t+\tau}(\Gamma)$ in the case of fixed
time ordering with 60 segments, to that with one segment corresponding to a linear scaling factor.
It can be seen that the value calculated with 60 weights,
$B^{(60)}_{t,t+\tau}(\Gamma)$, has a much broader distribution than $B^{(1)}_{t,t+\tau}(\Gamma)$
for a given value range of $\Omega_{t,t+\tau}(\Gamma;0)$, with a bias towards higher values for
$\Omega_{t,t+\tau}(\Gamma;0)$ near 0.
The skewed shape of the distribution near 0 is then compensated for further from
0, where a broader range of values will still result in the same prediction
(just with a different confidence), and hence the overall distribution of
$B^{(N)}_{t,t+\tau}(\Gamma)$ remains one that satisfies the FT.
Hence, it is evident that in the case of $B_{t,t+\tau}^{(60)}(\Gamma)$, the model is
recovering extra information beyond the value of $\Omega_{t,t+\tau}(\Gamma;0)$.
This is further evidenced by the distribution of the network weights, particularly for
longer trajectories, where samples of the dissipation function closer to the beginning
and end of the trajectory are more heavily weighted (see Figure \ref{fig:nn_weght_dist}).

It is also interesting to note that the FT was satisfied even in the case that
$N=1$ (see Figure \ref{fig:FT_random_weights}b).
Note that $B^{(1)}_{t,t+\tau}(\Gamma)$ is just
$\Omega_{t,t+\tau}(\Gamma;0)$ multiplied by a linear scaling factor ($B^{(1)}_{t,t+\tau}(\Gamma)
    = \frac{w}{\delta t} \Omega_{t,t+\tau}(\Gamma;0)$).
Hence, it is apparent (at least for the systems studied in this work) that the
assumption of linear correlation between the time-local and time-non-local
components of the dissipation function is sufficient for a good approximation of
the FT, even for very short trajectories.
That is, the correlations between $\Omega_{0,t}(\Gamma;0)$, $\Omega_{t,t+\tau}(\Gamma;0)$, and  $\Omega_{t+\tau,2t+\tau}(\Gamma;0)$ are linear to a good approximation when $t\rightarrow\infty$.

\subsection{Time correlations\label{sec:alpha_autocorrelation}}

\begin{figure}
    \centering
    \includegraphics[width=0.7\textwidth]{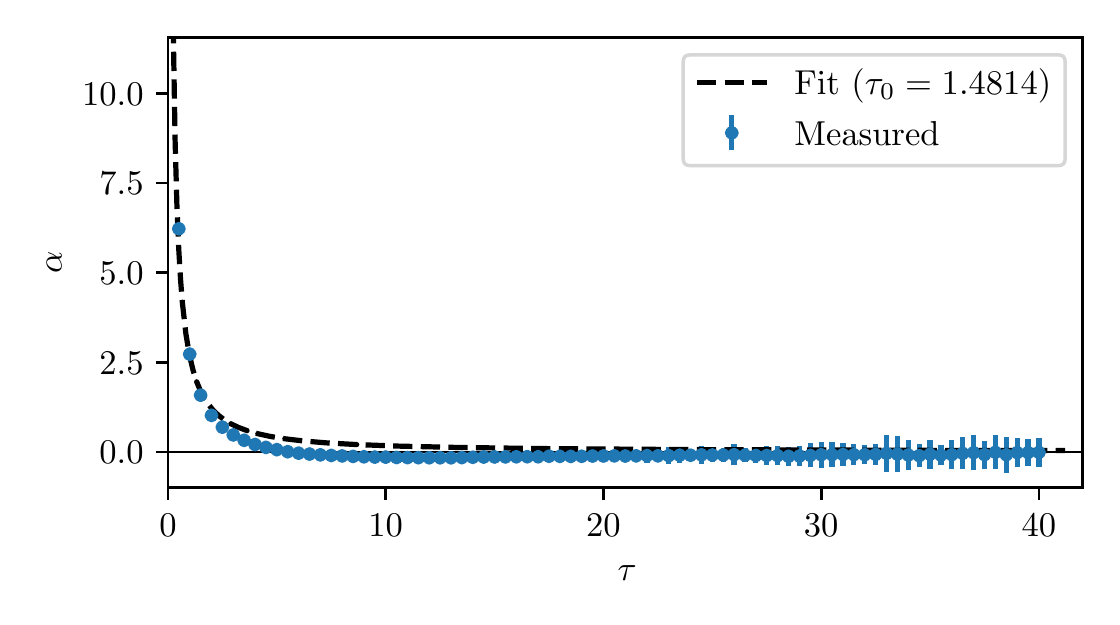}
    \caption{%
    Value of $\alpha$ required for the FT
    for $(1+\alpha)\Omega_{t,t+\tau}(\Gamma;0)$ to be best satisfied under the
    optical tweezers dynamics with $\rho=0.4$.
    $\alpha$ was fit to the slope of the FT considering only the inner 50\% of $A$
    values where statistics are relatively good,
    but the error bars represent the standard deviation across all available data points.
    The dashed line represents a fit in the form of Equation \ref{eqn:time-local-alpha}.
    \label{fig:alpha_vs_tau}
    }
\end{figure}

If only a single weight (equal to $1+\alpha$) is required to satisfy an FT,
it becomes simple with sufficient data to calculate this weight from the slope of
a plot of $\ln\left[p(\Omega_{t,t+\tau}(\Gamma;0)=A\tau)/p(\Omega_{t,t+\tau}(\Gamma;0)=-A\tau)\right]$ against $A\tau$.
As shown in Figure \ref{fig:alpha_vs_tau}, the required weight
as a function of $\tau$ is described reasonably well by the equation
\begin{equation}
    \alpha(\tau) = \frac{\tau_0\left(1-e^{-\tau/\tau_0}\right)}{\tau-\tau_0\left(1-e^{-\tau/\tau0}\right)},
    \label{eqn:time-local-alpha}
\end{equation}
where $\tau_0$ is the time constant of the decay in correlations,
as proposed for the space-local FT \cite{Michel2013,Talaei2012} based on the assumption of exponential
decay in spatial correlations (see Equation \ref{eqn:space_local_correlations}).
However, it is notable that for the dynamics modelled, the weights deviate from this form for a
range of mid-length trajectories.
This deviation (around $\tau=10$) includes values of $\alpha < 0$,
which cannot result from Equation \ref{eqn:time-local-alpha}.
It also corresponds closely to a minima in the autocorrelation function of the
instantaneous dissipation function, which similarly represents a region in which
the decay in time correlations does not follow an exponential relationship
(see Figure S3 in the supporting information).
Hence, it appears that derivations based on fitting the autocorrelation function of $\Omega(\Gamma;0)$
could result in a good approximation of $\alpha$ for steady state trajectories, $\Gamma_{t\rightarrow t+\tau}$,
provided the fit closely matches the autocorrelation data at a lag of $\tau$ (and regardless of whether
it is a good fit for other $\tau$ values).
That is, if only a specific range of $\tau$ values is of interest, then a fit to the autocorrelation
function of $\Omega(\Gamma;0)$ need only be accurate within that $\tau$ range in order to predict
values of $\alpha(\tau)$ which satisfy the FT.

\subsection{Calibration}
\begin{figure}
  \center\includegraphics[width=0.75\textwidth]{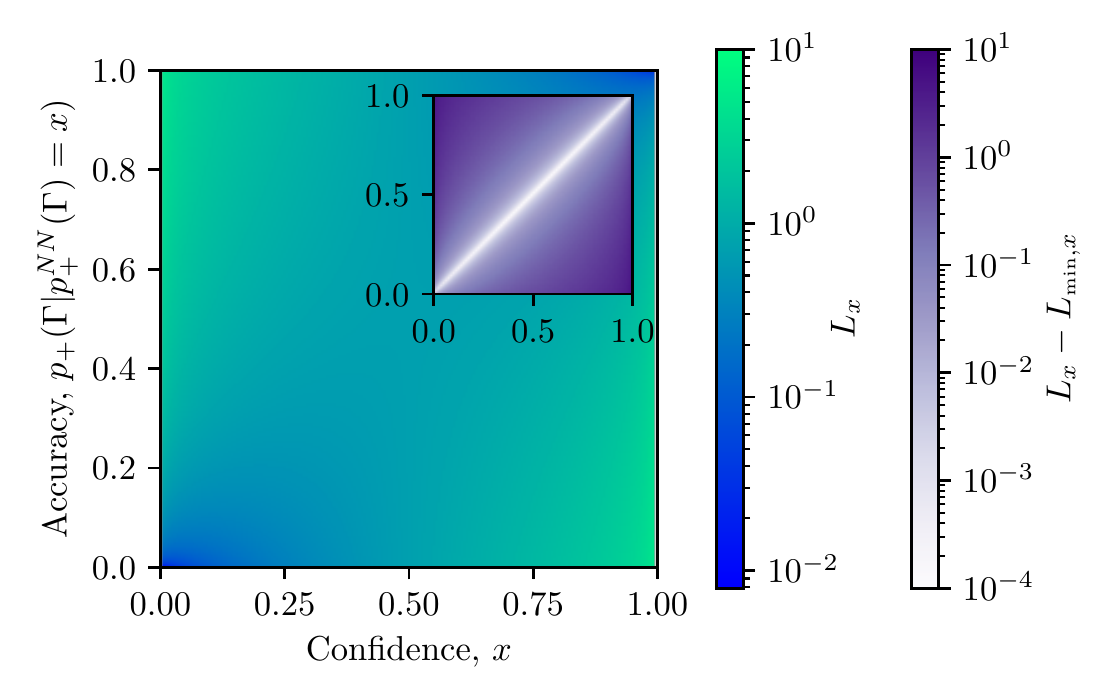}
  \caption{%
    Heat map of $L_x = \left<L(\Gamma)\right>_{p_+^{NN}(\Gamma)=x}$, the average value of the
    binary cross-entropy loss function, $L(\Gamma)$, for a given calibration specified
    by the confidence, $x$, and accuracy, $p_+(\Gamma|p_+^{NN}(\Gamma)=x)$.
    A well-calibrated predictor has a confidence equal to its accuracy.
    The inset shows the difference between $L_x$
    and its minimum value for a particular confidence, $L_{\text{min},x}$, clearly
    showing that reaching the global minimum of the binary cross-entropy loss function
    would result in a well-calibrated predictor.
  }
  \label{fig:loss_fn_heatmap}
\end{figure}

The models in this work were trained to minimise the binary cross-entropy loss
for a set of $N_t$ trajectories \cite{Seif2021,Goodfellow2016},
\begin{equation}
  L = -\frac{1}{N_t}\sum_{i=1}^{N_t}y_i\ln[p_+^{NN}(\Gamma_i)] +
  (1-y_i)\ln[1-p_+^{NN}(\Gamma_i)],
\end{equation}
where $y_i$ is 1 if the trajectory given by $\Gamma_i$ is a forward trajectory, or 0 if it has been
time-reversed.
Hence, for the set of trajectories assigned the particular probability
$p_+^{NN}(\Gamma)=x$ of being forward ones by the predictor, the average value
of the loss function is given by
\begin{equation}
  \left<L\right>_{p_+^{NN}(\Gamma)=x} =
      -p_+(\Gamma|p_+^{NN}(\Gamma)=x)\ln(x)
      -\left[1-p_+(\Gamma|p_+^{NN}(\Gamma)=x)\right]\ln(1-x),
\end{equation}
where $p_+(\Gamma|p_+^{NN}(\Gamma)=x)$ is the probability that a trajectory chosen
at random from the set was generated by the forward process as
opposed to the reverse process (i.e. the accuracy of the predictor for a given
confidence).
Figure \ref{fig:loss_fn_heatmap} plots $\left<L(\Gamma)\right>_{p_+^{NN}=x}$ as a heat map
against $p_+(\Gamma|p_+^{NN}(\Gamma)=x)$ and $x$, clearly showing
that the loss function is minimised when $p_+(\Gamma|p_+^{NN}(\Gamma)=x)=x$;
that is, when the accuracy matches the confidence.
Hence, any model that reaches the global minimum of the binary cross-entropy
loss function must be a well-calibrated predictor of time's arrow.

\subsection{Proof that a well-calibrated predictor satisfies a fluctuation theorem}
Since, from Equations \ref{eqn:sigmoid} and \ref{eqn:nn_probability},
\begin{eqnarray}
    p_+^{NN}(\Gamma) &=& \frac{1}{1+e^{-B^{(N)}_{t,t+\tau}(\Gamma)}}, \nonumber \\
        &=& \sigma\left(B^{(N)}_{t,t+\tau}(\Gamma)\right),
\end{eqnarray}
the probability that a trajectory with $B^{(N)}_{t,t+\tau}(\Gamma)=A\tau$ is a forward one as
opposed to a time-reversed one can be written based on the well-calibrated
condition as
\begin{eqnarray}
  p_+(\Gamma | B^{(N)}_{t,t+\tau}(\Gamma)=A\tau)
    &=& p_+(\Gamma | p_+^{NN}(\Gamma)=\sigma(A\tau)), \nonumber\\
    &=& \sigma(A\tau), \nonumber\\
    &=& \frac{1}{1+e^{-A\tau}}.
    \label{eq:well-calibrated}
\end{eqnarray}
This can be rearranged to give the ratio
\begin{equation}
  \ln\frac{p_+(\Gamma | B^{(N)}_{t,t+\tau}(\Gamma)=A\tau)}{p_-(\Gamma | B^{(N)}_{t,t+\tau}(\Gamma)=A\tau)}=A\tau,
\end{equation}
noting that $p_+ + p_- = 1$.
Since the predictor is well-calibrated, $B^{(N)}_{t,t+\tau}(\Gamma)$ must be odd
under time-reversal mapping, and therefore $p_-(\Gamma |
B^{(N)}_{t,t+\tau}(\Gamma)=A\tau)=p_+(\Gamma | B^{(N)}_{t,t+\tau}(\Gamma)=-A\tau)$.
Furthermore, odd time-reversal symmetry implies that
\begin{equation}
    p_+(\Gamma | B^{(N)}_{t,t+\tau}(\Gamma)=A\tau) =
    \frac{p(B^{(N)}_{t,t+\tau}(\Gamma)=A\tau)}
    {p(B^{(N)}_{t,t+\tau}(\Gamma)=A\tau)+p(B^{(N)}_{t,t+\tau}(\Gamma)=-A\tau)},
\end{equation}
and hence the form of the FT can be recovered:
\begin{equation}
  \ln\frac{p(B^{(N)}_{t,t+\tau}(\Gamma)=A\tau)}{p(B^{(N)}_{t,t+\tau}(\Gamma)=-A\tau)}=A\tau.
  \label{eq:FT_B}
\end{equation}
Thus, if Equation \ref{eq:well-calibrated} holds, the FT (Equation \ref{eq:FT_B}) will also be true.

It is therefore evident that any well-calibrated predictor of the direction
of time's arrow, will give a function which exactly satisfies the
FT.
Note that the probability assigned by the predictor need not be calculated
directly with the sigmoid function, but could instead be any combination of
non-linear functions that is able to produce a well-calibrated prediction.
In that case, the inverse sigmoid function of the resultant probability could be
used to obtain
\begin{equation}
  B^{(N)}_{t,t+\tau}(\Gamma) = -\ln\left(\frac{1}{p_+^{NN}(\Gamma)}-1\right).
\end{equation}
For ergodically consistent systems undergoing time-reversible dynamics, there is a
non-zero probability that a given trajectory could be a forward or time-reversed
one \cite{Evans2002}, meaning that this function will always be finite if $p_+^{NN}(\Gamma)$
exactly satisfies the well-calibrated condition.
However, in practice, small deviations from that condition, or numerical
inaccuracy caused by finite precision, may result in some infinite values of
$B^{(N)}_{t,t+\tau}(\Gamma)$, and hence the direct inclusion of the sigmoid
function in the model is preferable despite not being required.

\subsection{Uniqueness}
Note in the above derivation that there is no requirement that the predictor is the best possible one
for the system.
Based on the observations that multiple networks trained on the same
system produced different predictors, it appears that there are multiple
possible predictors which all satisfy an FT.
Indeed, the FT appears to be approximately satisfied for any
$B^{(N)}_{t,t+\tau}(\Gamma)$ where the weights are a good representation of the
required distribution.
It also appears to be satisfied when the weights are re-sampled from that
distribution for every trajectory, and when trajectories are randomly permuted
(i.e. the same trajectory could produce different values of
$B^{(N^*)}_{t,t+\tau}(\Gamma)$ depending on the particular permutation).
Since the satisfaction of the FT depends only on the well-calibrated condition,
$p_+(\Gamma | p_+^{NN}(\Gamma)=x) = x$, different partitioning of trajectories
achieved by different sets of weights (as observed in Figure
\ref{fig:dissipation_hist}) can all satisfy the FT.
Hence, predictors that satisfy the FT appear to be non-unique.

With some thought, this can be seen to apply even to transient systems beginning from a known distribution. Although the definition of the dissipation function is unique and given by Equation \ref{eqn:DFunc}, the predictors will depend on $t_2$.

\subsection{Implications of limited information and predictor complexity}
Consider the cases where only a linear scaling factor is required for $\Omega_{t,t+\tau}(\Gamma;0)$ to satisfy an FT to a good approximation; that is,
\begin{align}
  B_{t,t+\tau}^{(1)}(\Gamma) = \left<B_{t,t+\tau}^{(N^*)}(\Gamma)\right>
  &= (1+\alpha)\Omega_{t,t+\tau}(\Gamma;0),\nonumber\\
  &= \Omega_{t,t+\tau}(\Gamma;0)+\alpha\Omega_{t,t+\tau}(\Gamma;0).\nonumber\\
\end{align}
The fact that the FT is satisfied to a good approximation might seem at odds with the approximate SSFT,
\begin{equation}
  \ln\frac{p(\Omega_{t,t+\tau}(\Gamma;0)=A\tau)}{p(\Omega_{t,t+\tau}(\Gamma;0)=-A\tau)}
  = A\tau + \mathcal{O}(\tau_M).
\end{equation}
However, this apparent conflict can be reconciled by considering the case in the
approximate SSFT where
$\Omega_{t,t+\tau}(\Gamma;0)=0$.
In this case,
\begin{eqnarray}
  \ln\frac{p(\Omega_{t,t+\tau}(\Gamma;0)=+0)}{p(\Omega_{t,t+\tau}(\Gamma;0)=-0)}
    &=& 0 + \mathcal{O}(\tau_M), \nonumber\\
    &=& 0,
\end{eqnarray}
and hence the error term must have some dependence on
$\Omega_{t,t+\tau}(\Gamma;0)$.
As the value calculated by the model in the above cases differs from the
dissipation function only by a linear scaling factor, the corresponding FT is
\begin{equation}
  \ln\frac{p(\Omega_{t,t+\tau}(\Gamma;0)=A\tau)}{p(\Omega_{t,t+\tau}(\Gamma;0)=-A\tau)}
  = (1+\alpha) A\tau,
\end{equation}
which is exactly the form of the space-local FT when a linear correlation
between the local and non-local components of the dissipation function is
assumed \cite{Michel2013,Talaei2012}.
Hence, the FT satisfied by the model in the single weight case is expected to be
an approximate relation, although the approximation appears to be a good one for
all systems tested.
This, therefore, does not rule out the existence of higher order correlations.

It is notable that a simple scaling factor did not have as high a
predictive accuracy as that of a model in which consecutive segments of the
dissipation function were used as input (with time ordering preserved).
The root cause of this is difficult to state with certainty, but an analysis of
the weights showed that those towards the beginning and end of the trajectory
segment were consistently higher than those in the middle.
Furthermore, the shape of the weight plot (in the case of longer trajectories,
see Figure \ref{fig:nn_weght_dist}) was found to be qualitatively similar to that of a network
with the same structure (but excluding the sigmoid function) when
trained to predict the value of the time-local dissipation function for a longer
trajectory segment, $\Omega_{t-\Delta \tau,t+\tau+\Delta \tau}(\Gamma;0)$,
given, as in Figure \ref{fig:graphical_models}a, consecutive segments of
$\Omega_{t,t+\tau}(\Gamma;0)$.
This could indicate that the model is taking advantage of learned time
correlations to predict slightly longer trajectory segments, and thereby gain
more information to enable higher predictive accuracy \cite{Paneni2008}.
It could also be equivalently interpreted as the introduction of non-linear
correlations between the local and non-local components of the dissipation
function.

Regardless, it is clear that satisfaction of an FT (to a good approximation)
does not require full information about the trajectory, and the amount of
information available is more strongly correlated with the predictive accuracy
than with the accuracy of the FT.
Although perhaps surprising at first, a simple thought experiment shows that this
is an expected result.
Consider a transient system with trajectories of length $\tau$ beginning from a
known, equilibrium distribution.
Given reasonably closely spaced phase points (positions and momenta) along each
input trajectory, a sufficiently complex model could learn to effectively
integrate the equations of motion to some later time $\tau+\Delta\tau$ and
calculate the dissipation function $\Omega_{0,\tau+\Delta\tau}(\Gamma;0)$.
Both $\Omega_{0,\tau}(\Gamma;0)$ and $\Omega_{0,\tau+\Delta\tau}(\Gamma;0)$
satisfy the FT, but the latter would give better predictive accuracy due to the
extra information available.
In practice, though, if used to predict time's arrow for a set of fixed length
transient trajectories, such a strategy would not be beneficial.
This is because the non-equilibrium driving (and therefore the equations of motion)
have some time dependence which must be time-reversed for the reverse trajectories
(e.g. turning on the driving at the start of a forward trajectory versus turning
it off at the end of a reversed one).
The model would first have to determine which direction to integrate the equations of motion
before extending the trajectory, thereby requiring the prediction to already be made.
However, in a steady state where the non-equilibrium driving process is constant and $t\rightarrow\infty$,
both forward and time-reversed trajectories can simply be extended in both directions without issue.

Even with only knowledge of the instantaneous dissipation, and not the precise trajectory,
a sufficiently complex model could entirely learn the autocorrelation function of $\Omega(\Gamma;0)$,
and thereby reach a maximal predictive accuracy by capturing all correlations between $\Omega_{t,t+\tau}(\Gamma;0)$
and $\left(\Omega_{0,t}(\Gamma;0)+\Omega_{t+\tau,2t+\tau}(\Gamma;0)\right)$.
In such a case, it is clear that, similar to Equation \ref{eqn:local_ft_splitting}, the dissipation
function from time $0$ to time $2t+\tau$, which satisfies the Evans-Searles FT, can be expressed as
\begin{equation}
    \Omega_{0,2t+\tau}(\Gamma;0) = B(\Omega_{t,t+\tau}(\Gamma;0)) + \xi(\Gamma),
\end{equation}
where $B(\Omega_{t,t+\tau}(\Gamma;0))$ represents the value computed by the model
containing all correlations, and $\xi(\Gamma)$ is the remaining uncorrelated component
which will depend on the particular trajectory.
As  $\Omega_{t,t+\tau}(\Gamma;0)$ is uncorrelated with $\xi(\Gamma)$, $B(\Omega_{t,t+\tau}(\Gamma;0))$
must also be uncorrelated, and therefore, since $\Omega_{0,2t+\tau}(\Gamma;0)$ is odd under
time reversal mapping, so must be both $B$ and $\xi$.
Hence, it follows from the functional FT \cite{Michel2013,Talaei2012,Ayton2000,Searles2013,Evans2016}
(Equation \ref{eqn:functional_ft}) that
\begin{align}
    \ln\frac{p(\Omega_{t,t+\tau}(\Gamma;0)=A\tau)}{p(\Omega_{t,t+\tau}(\Gamma;0)=-A\tau)}
        &= -\ln\left<e^{-\Omega_{0,2t+\tau}(\Gamma;0)}\right>_{\Omega_{t,t+\tau}(\Gamma;0)=A\tau}, \nonumber\\
        &= -\ln\left<e^{-B(\Omega_{t,t+\tau}(\Gamma;0))}e^{-\xi(\Gamma)}\right>_{\Omega_{t,t+\tau}(\Gamma;0)=A\tau}, \nonumber\\
        &= -\ln\left<e^{-B(A\tau)}
            e^{-\xi(\Gamma)}\right>_{\Omega_{t,t+\tau}(\Gamma;0)=A\tau}, \nonumber\\
        &= B(A\tau) - \ln\left<e^{-\xi(\Gamma)}\right>, \nonumber\\
        &= B(A\tau),
    \label{eqn:exact-local-FT}
\end{align}
where the final line is obtained by noting that since $B$ is an odd function,
the term involving the ensemble average must be zero in the $A=0$ case,
and must therefore always be zero as it is uncorrelated with $A$.
This local FT is an exact equality which depends only on time-local information and
information that can be obtained with knowledge of the autocorrelation function of
$\Omega(\Gamma;0)$, and does not depend on any uncorrelated dissipation.
Note that although it is expressed here in terms of time-locality, Equation
\ref{eqn:exact-local-FT} applies to any local dissipation function that is
a component of a global dissipation function which satisfies a FT, including
the steady state space-local dissipation function, $\Omega_{t,t+\tau}^{l}(\Gamma;0)$,
which is often the easiest to measure experimentally.

\subsection{Applications}
Finally, it is apparent that optimising the binary cross-entropy loss of a
predictor of time's arrow presents a method with which a
scaling factor, $\alpha$, could be calculated for a system of interest and thereby
result in a quantity which satisfies a FT to a good approximation.
As $\alpha$ depends on both the dynamics and the length of the trajectory
segments, calculating it via the slope of a plot of
$\ln[p(\Omega_{t,t+\tau}(\Gamma;0)=A\tau)/p(\Omega_{t,t+\tau}(\Gamma;0)=-A\tau)]$ versus $A\tau$
may be infeasible in many cases due to the large number of trajectories
required, and similar problems may be encountered in attempting to use the
autocorrelation function of $\Omega(\Gamma_t;0)$ as in Section \ref{sec:alpha_autocorrelation}
or Equation \ref{eqn:exact-local-FT} if correlations decay too slowly.
However, testing showed that the logistic regression model converged to
a predictor which closely satisfied an FT from very few input trajectories in some
cases.
For example, training on data from the SLLOD dynamics with $\dot\gamma=0.5$ and
$\tau=0.01$, only 64 input trajectories (32 forward, 32 reverse) were required
to calculate a converged value of $\alpha$.
As would be reasonably expected, this required significantly more training
epochs than when training on larger data sets (the model needed to `see' each
trajectory more times).
However, a converged $\alpha$ value was achieved both in the case of a
single-weight model ($N = 1$), and one with $N = 10$ where
the samples of the dissipation function were randomly permuted.
In the latter case, $\alpha$ was determined by the average of the weights.

One potential problem when calculating $\alpha$ in this manner is that verification
by plotting the FT relation requires just as much data as determining it by the
slope of a plot of
$\ln[p(\Omega_{t,t+\tau}(\Gamma;0)=A\tau)/p(\Omega_{t,t+\tau}(\Gamma;0)=-A\tau)]$ versus $A\tau$.
However, this could be avoided by simply increasing the number of input
trajectories until the calculated $\alpha$ converges.
Note that exact convergence can be difficult to achieve even with a large
data set, as there is room for some variance without major impact on the overall
predictive accuracy (see Figure \ref{fig:loss_fn_heatmap}).
However, retraining multiple times on different subsets of the data and
averaging over the calculated $\alpha$ values was found to give a good
approximation, and a modified loss function that strongly rewards the
well-calibrated condition may also prove beneficial.
Such a procedure may have applications in predicting the probability of rare events,
or in calculating phase variable averages at steady states, although we leave this for future work.

\section{Conclusion}

Machine learning models were trained to predict whether a given deterministic,
nonequilibrium steady state trajectory was one that progressed forward or
backward in time.
It was found that a value which satisfies a fluctuation theorem can be
calculated from the model's output provided the model converged to a
well-calibrated predictor (a condition shown to be automatically satisfied if
the binary cross-entropy loss is exactly minimised).
The FT was satisfied even for very short trajectories where the approximate
time-local steady state FT derived from theory is not valid,
and this result was verified for a set of common nonequilibrium dynamics at
various distances from equilibrium.

It was shown that even a simple linear scaling factor was sufficient to give a good
approximation of the FT when applied to the time-local dissipation function
defined with respect to the initial (known) phase space distribution.
This is equivalent to previously derived space-local FTs.
It was also observed that a more complex model was able to capture
higher order correlations between the time-local and non-local components of the
dissipation function, and thereby produce a higher predictive accuracy.
Hence, it was demonstrated that although the amount of information available about
the trajectory has some correlation with the predictive accuracy, it had no
noticeable effect on the accuracy with which the FT was satisfied for the
systems tested.
Furthermore, it was demonstrated that an exact local FT can be written which
depends only on local information and the correlation between that information
and the unknown non-local component.
Finally, a method was proposed for calculating the scaling factor required to
approximately satisfy the FT in the case that only a small data set is available.

\section*{Acknowledgements}

The authors thank the Australian Research Council for its support for this project through the Discovery program (FL190100080). We acknowledge access to computational resources provided by the Pawsey Supercomputing Centre with funding from the Australian Government and the government of Western Australia, and the National Computational Infrastructure (NCI Australia), an NCRIS enabled capability supported by the Australian Government. We also acknowledge support from the Queensland Cyber Infrastructure Foundation (QCIF). Finally, we would like to thank Dr. Shern Tee for his insightful comments and suggestions.

\vspace{0.2cm}

\let\doi\relax

\bibliography{references}



\section*{Supplementary material}

# Supporting Information: Machine learning a time-local fluctuation theorem for nonequilibrium steady states

Stephen Sanderson[1], Charlotte F. Petersen[1,2], and Debra J. Searles[1,3,*]

[1]Australian Institute for Bioengineering and Nanotechnology, The University of Queensland, Brisbane, QLD, 4072, Australia
[2]School of Chemistry, University of Melbourne, Melbourne, Victoria, 3010, Australia
[3]School of Chemistry and Molecular Biosciences, The University of Queensland, Brisbane, QLD, 4072, Australia
[*]d.bernhardt@uq.edu.au

## 1 Derivation of $p_+(\Gamma_{0\to\tau})$ from the fluctuation theorem

Consider configurations sampled from an initial equilibrium state that evolve with a given protocol to configurations that would sample a new equilibrium state, if allowed to relax. Let the free energy difference in these equilibrium states be $\Delta F$. The Crooks fluctuation theorem states that

$$\frac{p_F(W_F = A)}{p_R(W_R = -A)} = e^{\beta(A-\Delta F)}, \tag{S1}$$

where $p_F(W_F = A)$ is the probability of observing the value, $A$, of the work when driving a system from the initial equilibrium ensemble to the final configuration and the probability, $p_R(W_R = -A)$, is then the probability of observing the value of work, $-A$, when driving the system from the equilibrium distribution of the final configuration back to the initial configuration, under the reverse protocol. This relationship can be obtained by considering the ratio of probabilities of observing forward trajectories within a phase volume, $\delta\Gamma$, about point $\Gamma_0$, sampled from the initial equilibrium ensemble; compared to the probability of observing 'reverse' trajectories in the phase volume it would evolve to, $\Gamma_\tau$, subject to a time reversal mapping ($\Gamma_0^* = M^T(\Gamma_\tau)$) and evolved using the reverse protocol. This gives, [1, 2]

$$\frac{p_F(\delta\Gamma_0)}{p_R(\delta\Gamma_0^*)} = \frac{p_F(\Gamma_{0\to\tau})}{p_R(\Gamma_{0\to\tau}^*)} = e^{\beta(W(\Gamma_{0\to\tau})-\Delta F)}. \tag{S2}$$

Consider a mix of trajectories in which an equal number were generated by the forward process and reverse process, with the reverse trajectories then played backwards. The probability that a given trajectory was generated by the forward one rather than the reverse one is given by

$$p_+(\Gamma_{0\to\tau}) = \frac{p_F(\Gamma_{0\to\tau})}{p_F(\Gamma_{0\to\tau}) + p_R(\Gamma_{\tau\to 0})}, \tag{S3}$$

where $p_R(\Gamma_{\tau\to 0}) = p_R(\Gamma_{0\to\tau}^*)$ is the probability that the trajectory was generated by the reverse process, and is played backwards. Then,

$$p_+(\Gamma_{0\to\tau}) = \frac{p_F(\Gamma_{0\to\tau})}{p_F(\Gamma_{0\to\tau}) + p_R(\Gamma_{0\to\tau}^*)}, \tag{S4}$$

$$= \frac{p_F(\Gamma_{0\to\tau})}{p_F(\Gamma_{0\to\tau}) + p_F(\Gamma_{0\to\tau})e^{-\beta(W(\Gamma_{0\to\tau})-\Delta F)}}, \tag{S5}$$

$$= \frac{1}{1 + e^{-\beta(W(\Gamma_{0\to\tau})-\Delta F)}}. \tag{S6}$$

Using the same logic, an equivalent expression can be derived from the Evans-Searles fluctuation theorem. [3] That is, the arguments leading to the Evans-Searles fluctuation theorem give:

$$\frac{p(\delta\Gamma_0)}{p(\delta\Gamma_0^*)} = \frac{p(\Gamma_{0\to\tau})}{p(\Gamma_{0\to\tau}^*)} = e^{\Omega_{0,\tau}(\Gamma;0)}. \tag{S7}$$

Then, assuming an equal mix of forward and time-reversed trajectories, this expression can then be rearranged as before to give the probability that a trajectory is a forward one as

$$p_+(\Gamma_{0\to\tau}) = \frac{1}{1+e^{-\Omega_{0,\tau}(\Gamma;0)}}. \tag{S8}$$

## 2 Illustration of typical forward and reverse trajectories

Figure S1 represents the relationships between the time-evolution of the dissipation function along trajectories that are played in the forward and reverse directions, and the likelihood of them being observed for a typical driven system, initially at equilibrium and reaching a steady state. The dissipation function would be expected to obey a fluctuation theorem as presented in Equation 1 of the main text.

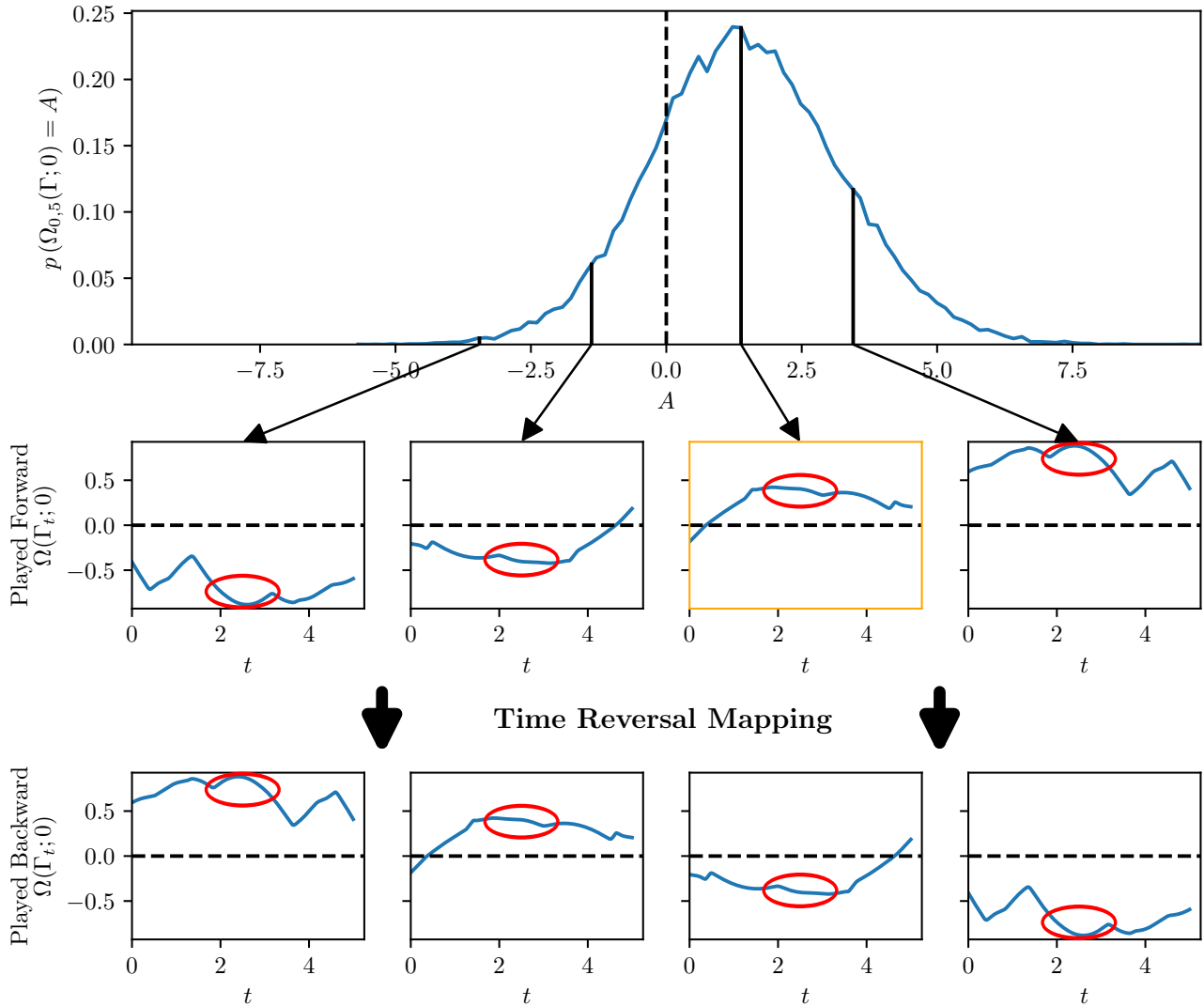

**Figure S1.** Schematic diagram illustrating the probability distribution of the dissipation of trajectories $\Gamma_{0\to 5}$, typical trajectories generated when a driving force is applied at $t = 0$, and the trajectories that would be observed if the original set were played backwards. The four representative plots of the instantaneous dissipation function against time are arranged by the trajectories' value of $A$, as indicated. The bottom four plots illustrate what these same trajectories would look like if played in reverse. The trace highlighted in orange is a trajectory that would be typical of the forward process. The circled regions represent portions of steady state trajectories. This diagram was generated using transient trajectories from the optical tweezers experiment with a fluid number density of $\rho = 0.3$, $k_{\text{trap}} = 1.0$, $v_{\text{trap}} = 0.5$, and $k_B T = 1.0$.

## 3 Benchmarking for deterministic transient systems

Linear regression models were trained on transient trajectories obtained using non-equilibrium simulations of the optical trap system beginning from an equilibrium state. These transient trajectories all began from different initial conditions taken at regular time intervals from the equilibrium system. While various input data was trialled, the only input that mattered under this simple model was the position of the trapped particle, *x*, in the direction that the trap was moving, which is directly related to the instantaneous dissipation function for these dynamics. Samples of *x* were taken at regular time intervals. Figure S2a shows that the network trained best when given enough samples of the trajectory to have near-complete information while keeping the number of free parameters to a minimum. As this figure was generated using the position of the particle relative to the *initial* position of the trap, a bias was required to achieve good results. The fit of the model's output for the best-performing sampling period was then compared in Figure S2b to the relation expected based on the Evans-Searles fluctuation theorem (Eqn. S8), and found to be in good agreement.

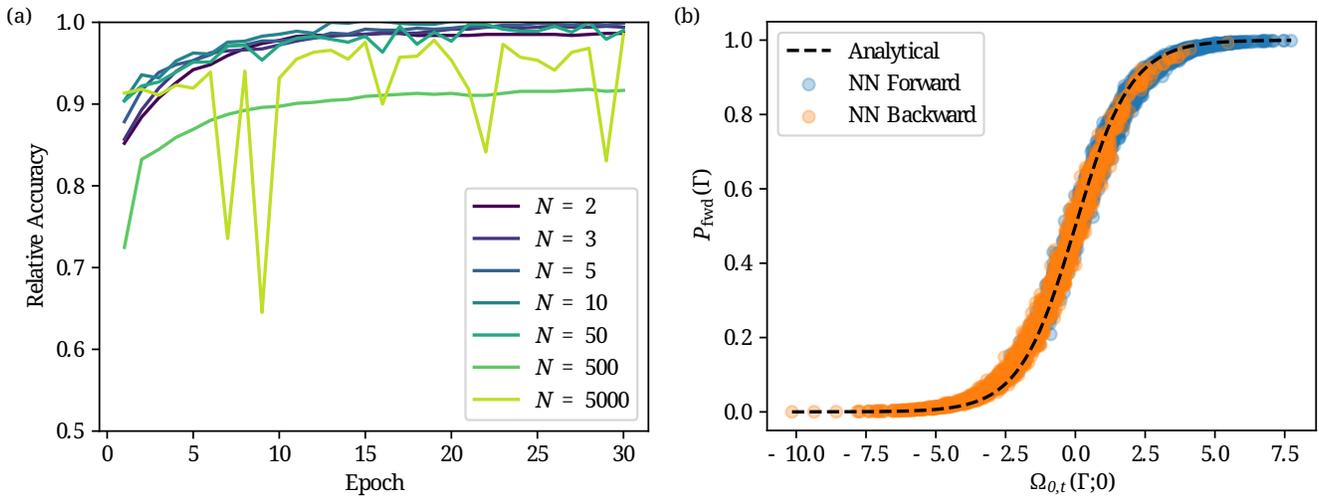

**Figure S2.** (a) Training accuracy for *N* instantaneous samples of the trapped particle's position along each transient deterministic trajectory, and (b) model predictions plotted in comparison to Equation S8 for $N = 10$, where blue circles represent trajectories that were played forwards, and orange those that were played backwards. Trajectories were generated with the transient optical tweezers dynamics (beginning from equilibrium) for $\rho = 0.3$, $v_{\text{trap}} = 0.5$, $k_{\text{trap}} = 1.0$, and $k_B T = 1.0$.

## 4 Autocorrelation of the instantaneous dissipation function

Figure S3 shows the autocorrelation function of the instantaneous dissipation function at steady state for the same dynamics used to plot Figure 5c in the main text. Note the deviation from an exponential decay near $\Delta \tau = 10$ corresponds closely with the region of Figure 5c in which the assumption of an exponential decay in correlations does not produce a good fit for $\alpha$.

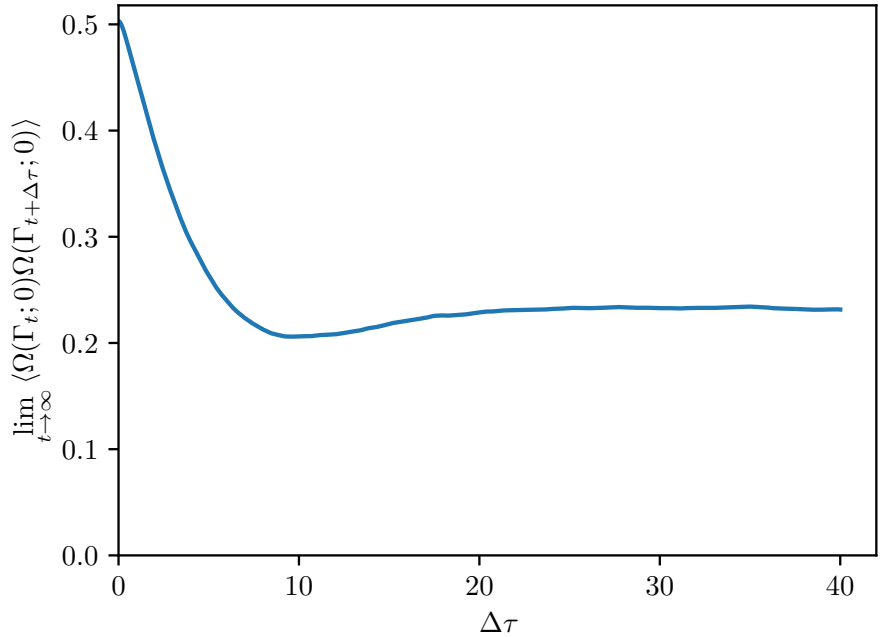

**Figure S3.** Autocorrelation function of the instantaneous dissipation function, $\Omega(\Gamma_t; 0)$, calculated from 100,000 repetitions of the optical tweezers dynamics with $\rho = 0.4$, $v_{\text{trap}} = 0.5$, $k_{\text{trap}} = 1.0$, and $k_B T = 1.0$.



\end{document}